\documentclass{aa}

\usepackage{graphicx}
\usepackage{longtable}
\usepackage{pdflscape}
\usepackage[normalem]{ulem}
\usepackage{txfonts}
\usepackage{natbib}
\begin{document}

   \title{Complex cyanides as chemical clocks in hot cores}

%   \subtitle{with ALMA}

   \author{V. Allen\inst{1,2}
          \and
          F. F. S. van der Tak\inst{1,2}\fnmsep
          \and C. Walsh\inst{3}
          }

   \institute{Kapteyn Astronomical Institute, University of Groningen,
              the Netherlands\\
              \email{allen@astro.rug.nl, vdtak@sron.nl}
             \and SRON, Groningen, the Netherlands
             \and School of Physics and Astronomy, University of Leeds, Leeds, LS2 9JT, UK
             }

   \date{Received 2017; accepted 2017}

% \abstract{}{}{}{}{} 
% 5 {} token are mandatory
 
  \abstract
   {In the high-mass star-forming region G35.20-0.74N, small scale ($\sim$800~AU) chemical segregation has been observed in which complex organic molecules containing the CN group are located in a small location (toward continuum peak B3) within an apparently coherently rotating structure.}
   {We aim to determine the physical origin of the large abundance difference ($\sim$~4 orders of magnitude) in complex cyanides within G35.20-0.74 B, and we explore variations in age, gas/dust temperature, and gas density.}
   {We performed gas-grain astrochemical modeling experiments with exponentially increasing (coupled) gas and dust temperature rising from 10 to 500~K at constant H$_2$ densities of 10$^7$~cm$^{-3}$, 10$^8$~cm$^{-3}$, and 10$^9$~cm$^{-3}$. We tested the effect of varying the initial ice composition, cosmic-ray ionization rate (1.3$\times$10$^{-17}$ s$^{-1}$, 1$\times$10$^{-16}$ s$^{-1}$, and 6$\times$10$^{-16}$ s$^{-1}$), warm-up time (over 50, 200, and 1000~kyr), and initial (10, 15, and 25~K) and final temperatures (300 and 500~K).}
   {Varying the initial ice compositions within the observed and expected ranges does not noticeably affect the modeled abundances indicating that the chemical make-up of hot cores is determined in the warm-up stage.  Complex cyanides vinyl and ethyl cyanide (CH$_2$CHCN and C$_2$H$_5$CN, respectively) cannot be produced in abundances (versus H$_2$) greater than 5$\times10^{-10}$ for CH$_2$CHCN and 2$\times10^{-10}$ for C$_2$H$_5$CN with a fast warm-up time (52~kyr), while the lower limit for the observed abundance of C$_2$H$_5$CN toward source B3 is 3.4$\times10^{-10}$.  Complex cyanide abundances are reduced at higher initial temperatures and increased at higher cosmic-ray ionization rates. Reaction-diffusion competition is necessary to reproduce observed abundances of oxygen-bearing species in our model.}
   {Within the context of this model, reproducing the observed abundances toward G35.20-0.74 Core B3 requires a fast warm-up at a high cosmic-ray ionization rate ($\sim$1$\times$10$^{-16}$~s$^{-1}$) at a high gas density ($>$10$^9$~cm$^{-3}$). The abundances observed at the other positions in G35.20-0.74N also require a fast warm-up but allow lower gas densities ($\sim$10$^8$~cm$^{-3}$) and cosmic-ray ionization rates ($\sim$1$\times$10$^{-17}$~s$^{-1}$). In general, we find that the abundance of ethyl cyanide in particular is maximized in models with a low initial temperature, a high cosmic-ray ionization rate, a long warm-up time ($>$~200~kyr), and a lower gas density (tested down to 10$^7$~cm$^{-3}$).  G35.20-0.74 source B3 only needs to be $\sim$2000 years older than B1/B2 for the observed chemical difference to be present, which maintains the possibility that G35.20-0.74 B contains a Keplerian disk.}

   \keywords{stars: massive -- ISM: individual objects: G35.20-0.74N -- astrochemistry -- ISM : molecules}

   \maketitle
%
%________________________________________________________________

\section{Introduction}

  In high-mass star formation, the hot molecular core (HMC) stage is marked by high abundances of complex organic molecules (COMs), molecular species containing at least six atoms including carbon and hydrogen \citep{Herbst2009}, and emitting from a warm (100-500~K), dense (n$_H$ $>$10$^7$~cm$^{-3}$), and compact ($<$0.05~pc) region.  The physical nature of this type of region, whether a disk, an outflow cavity, or an envelope, is currently unknown.  The COMs seen in hot cores are expected to be abundantly produced in the ice mantles hosted on dust grains around the forming star and released into the gas phase upon warming. Also, COMs can be produced through endothermic reactions in warm gas.  The HMC stage is not expected to last more than 10$^5$ years, as COMs are dissociated in the expanding HII region around a young high-mass star.  As a short-lived stage with specific physical parameters, the HMC is an ideal source for studying the process of high-mass star formation and by tracing the distribution of specific molecular species, we can learn more about the physical and chemical structure of these young objects.
\par Chemical segregation has been observed in several different star-forming regions on scales from 1000-8000~AU, most famously in Orion KL where \citet{Blake1987} observed that the hot core has a much higher abundance of N-bearing species than the compact ridge and surrounding sources.  To explain this, \citet{PaolaOrion1993} modeled shells of gas collapsing toward the nearby object IRc2, which are halted and heated up showing different chemical compositions (see \citet{Siyifeng2015} and \citet{Crockett2015} for recent work on Orion KL).  A difference in chemical composition has also been seen between W3(OH) and W3(H$_2$O) \citep{Wyrowski1999} where the latter is a strong N-bearing source with various complex organics, but the former only contains a handful of O-bearing species.  AFGL2591 VLA 3 is another source \citep{Izaskun2012} where such chemical segregation has been observed on a scale of a few thousand AU, which was explained using models of concentric shells with different temperatures and amounts of extinction.

   \begin{figure}[h]
   \centering
      \includegraphics[width=\hsize]{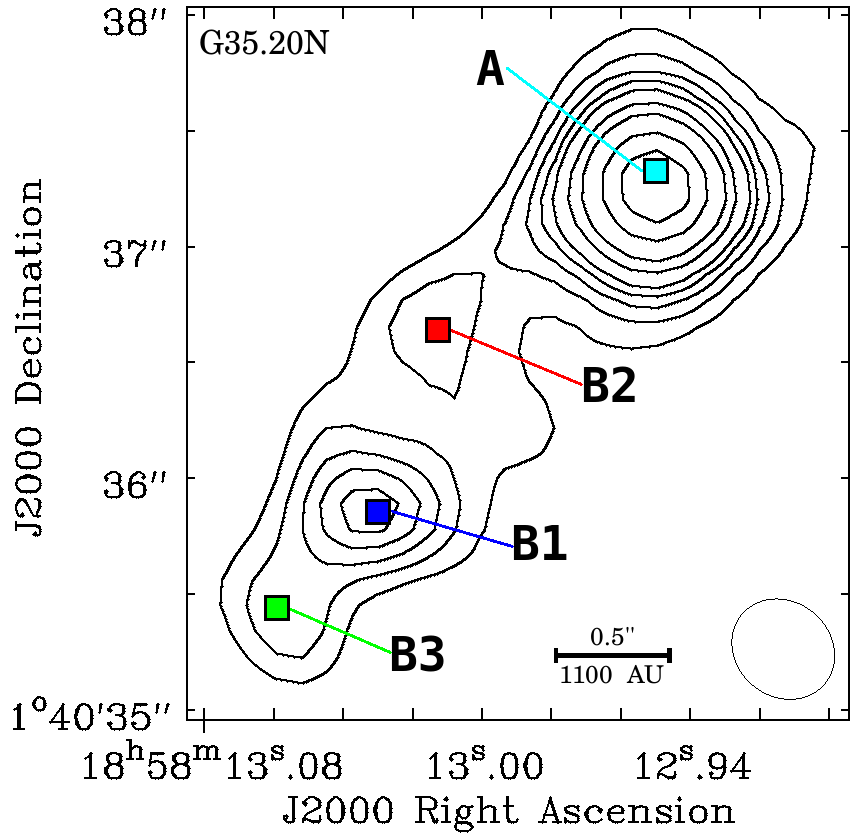}
      \caption{Image of the 870 $\mu$m continuum emission from Cycle 0 ALMA observations of G35.20.  The continuum peaks are labeled in order of intensity (i.e., peak B1 has the highest and peak B3 the lowest continuum intensity). Contour levels are 0.03, 0.042, 0.055, 0.067, 0.08, 0.10, 0.13, 0.18, and 0.23 Jy/beam ($\sigma$ $=$ 1.8 mJy/beam).  The pixel-sized colored squares denote each of the spectral extraction points  (from \citet{Allen}).}
         \label{ContPts3520}
   \end{figure}

\par This paper follows our previous study \citep{Allen} of G35.20-0.74N (G35.20), a high-mass star-forming region containing several high-mass protostars at a distance of 2.19~kpc with a bolometric luminosity of $3.0\times10^4$ L$_{\sun}$ \citep{Alvaro2014}. G35.20 was shown to be a strong Keplerian disk candidate based on position-velocity diagrams for several species and the fit of the velocity field to a Keplerian disk model \citep{Alvaro2013}.  In this source, we observed a segregation in Core B between complex N-bearing species, especially cyanides (those containing the CN group), and other  COMs on a scale of less than 1000~AU within an apparently coherent source presenting a potential signature of Keplerian rotation.  Within Core B (shown in Figure~\ref{ContPts3520}) there is a higher abundance (generally 1-2 orders of magnitude) of almost all observed species to the southeast (at continuum peak B3), and additionally, the nitrogen-bearing species abundance drops quickly when proceeding to the northwest (to continuum peaks B1 and B2 about 1000 and 2000 AU from source B3, respectively) where most complex N-bearing species (especially those with the CN group) are no longer detected.  This is especially prominent in ethyl- and vinyl cyanide (C$_2$H$_5$CN and CH$_2$CHCN) and in vibrationally excited states and isotopologs of methyl cyanide (CH$_3$CN) and cyanoacetylene (HC$_3$N). We also model the observed abundances from Core A for comparision, as it is not part of the potential Core B disk system, but has a similar chemical composition to source B3 with high abundances of cyanides and oxygen-bearing species.
\par We expect G35.20 source B3 to be a high-mass source as a high kinetic temperature is observed toward peak B3 ($\sim$300~K) compared to peak B1 and peak B2 (160 and 120~K, respectively). Alongside this high temperature, the deuterium fraction is very high toward source B3 (13$\%$ for CH$_3$CN) implying that it has only recently heated up, releasing deuterium enriched ices into the gas phase. There is also a cluster of OH masers toward peak B3 \citep{OHmasers1999}.  At the outer radius of this disk candidate, the rotation period is between 9700 and 11100~years, which is fast enough that such a difference in chemistry should not be present because of the mixing of gas.  In this work, we use chemical modeling to investigate a cause for chemical segregation between complex cyanides and other species related to age, temperature, warm-up time, or gas density.

\section{Chemical model}

\subsection{Model setup}
\label{chemModel}

We used a large gas-grain chemical network (668 species, over 8000 reactions) in which the gas-phase reactions are from the UMIST Database for Astrochemistry \citep{McElroy2013} known as Rate12\footnote{http://www.udfa.net/}, and the grain surface and gas-grain interactions are extracted from the Ohio State University (OSU) network (detailed description in \citet{Walsh2014}).  Our network includes the following reaction types: two-body gas-phase reactions, direct cosmic-ray ionization, cosmic-ray-induced photoreactions, photoreactions, cation-grain recombination, adsorption onto grains, thermal desorption, photodesorption, grain-surface cosmic-ray-induced photoreactions, grain-surface photoreactions, two-body grain-surface reactions, and reactive desorption.
\par In this model, the thermal desorption rate depends on the binding energy of the species ($\mathrm{E_{bind,A}}$) and the number density of that species on the grain surface ($n\mathrm{_s(A)}$). If there are less than two monolayers, then the following first order rate is used: $f\mathrm{_{thermal,A}} = k\mathrm{_{evap,A}}n\mathrm{_s(A)}$ \citep{Cuppen}, where $k\mathrm{_{evap,A}=\nu~exp}^{-\frac{E\mathrm{_{bind,A}}}{kT}}$ and $\mathrm{\nu}$ is the characteristic attempt frequency.  Once there are more than two monolayers, then the following zeroth-order approximation: $f\mathrm{_{thermal,A} = }k\mathrm{_{evap,A}}N\mathrm{_{act}\chi_A}N\mathrm{_s\sigma_g}n\mathrm{_{grain}}$, where $N\mathrm{_{act}}$ is the number of active monolayers, $\mathrm{\chi_A}$ is the fractional abundance of species A, and $N\mathrm{_s\sigma_g}n\mathrm{_{grain}}$ is the number of available surface sites per unit volume. Further details about this chemical code can be found in \citet{Droz2014}, \citet{Walsh2014}, and \citet{Walsh2015}. Reaction-diffusion competition is included.  
\par The model considers a single embedded (A$_V$=10) point source at a constant gas density that is warming up over time.  The relatively high extinction means that the only source of ionization and photodissociation in the model is cosmic rays.  A low cosmic-ray ionization rate of $1.3\times10^{-17}$ s$^{-1}$ was used in the fiducial model. Higher cosmic-ray ionization rates are explored in test cases (\S~\ref{RDC}-\ref{CRi}).  

 \begin{table}[!ht]
  \centering
  \caption{Initial ice composition vs. H$_2$O ice.  The initial H$_2$ abundance for all models is 50$\%$ of the total material.  The H$_2$O ice abundances vs. the total composition are 5$\times$10$^{-6}$ for IC 1, 5$\times$10$^{-5}$ for IC 3, and 10$^{-5}$ for IC 2, 4, and 5.}
  \label{ICTable}
  \begin{tabular}{cccccc}

  \hline\hline
  Species & IC 1 & IC 2 & IC 3 & IC 4 & IC 5 \\
  \hline
  CO (ice) & 10$\%$ & 5$\%$ & 10$\%$ & 8$\%$ & 17$\%$ \\
  CO$_2$ (ice) & 10$\%$ & 15$\%$ & 10$\%$ & 13$\%$ & 23$\%$ \\
  NH$_3$ (ice) & 5$\%$ & 2$\%$ & 5$\%$ & 15$\%$ & 15$\%$ \\
  CH$_3$OH (ice) & 5$\%$ & 5$\%$ & 5$\%$ & 10$\%$ & 4$\%$ \\
  HCOOH (ice) & 10$\%$ & 5$\%$ & 10$\%$ & 7$\%$ & 1$\%$ \\
  CH$_4$ (ice) & 5$\%$ & 1$\%$ & 5$\%$ & 1.5$\%$ & 1.5$\%$ \\
  H$_2$CO (ice) & 10$\%$ & 2$\%$ & 10$\%$ & 3.5$\%$ & 2$\%$ \\
  \hline
  \end{tabular}
  \end{table}

\begin{figure*}[!htb]
   \centering
      \includegraphics[width=\hsize]{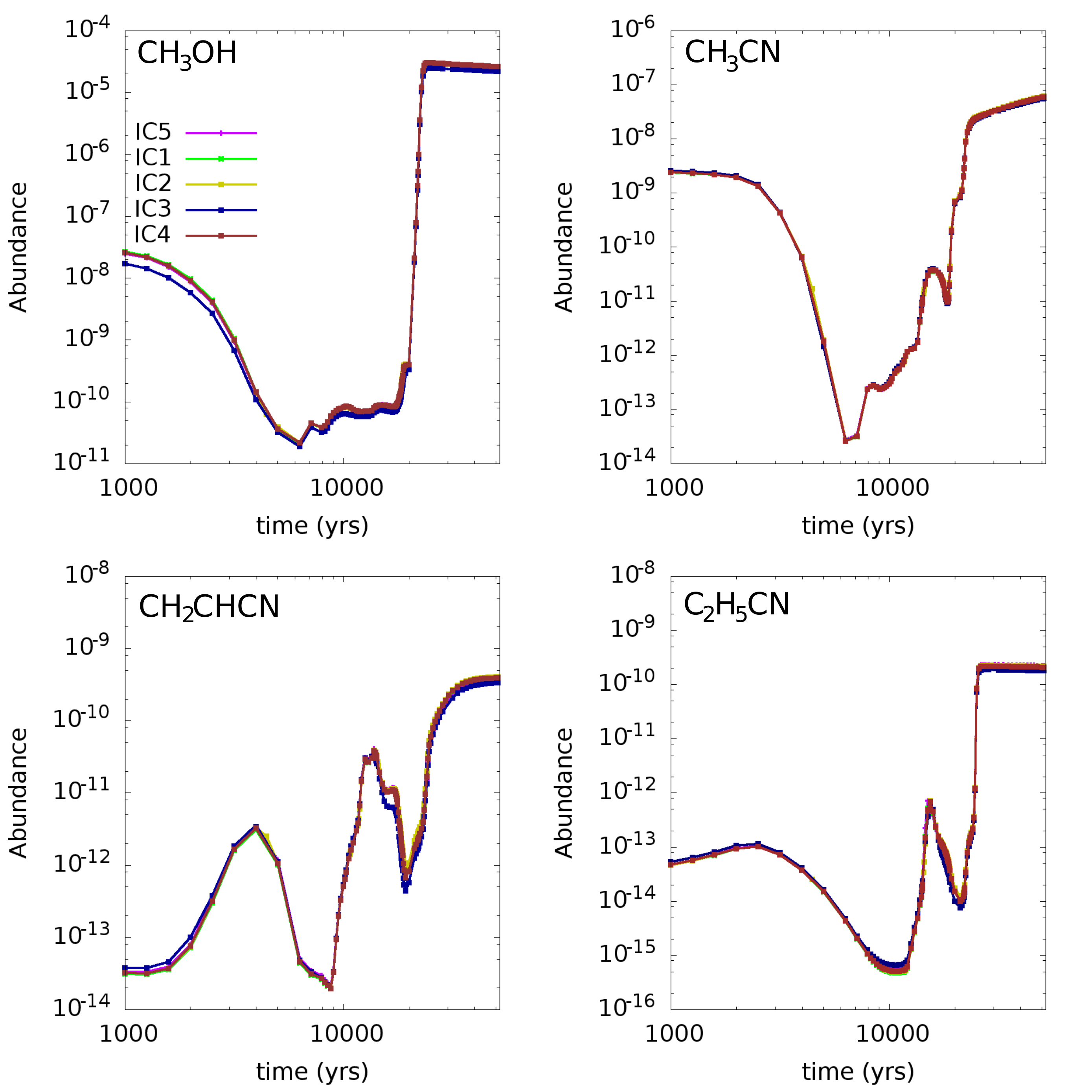}
      \caption{Fractional abundance vs. H$_2$ for one species with different initial conditions (ICs 1-5) for a gas density of 10$^7$ cm$^{-3}$ and a fast warm-up time of 50 kyr. Results are similar for other warm-up times and densities.}
         \label{ICs}
   \end{figure*}

\subsection{Initial conditions}
\label{ICsec}
We start a warm-up phase at the end of a theoretical collapse phase that results in a constant H$_2$ density of n = 10$^7$, 10$^8$, or 10$^9$ cm$^{-3}$ with enhanced ice abundances of several species (see Table~\ref{ICTable}).  The gas density of Core B is expected to be 10$^7$-10$^8$ cm$^{-3}$ and Core A is expected to have a density of 10$^9$ cm$^{-3}$ from \citet{Alvaro2014}.  The warm-up phases start with the initial conditions (IC) outlined in Table~\ref{ICTable} and the (coupled) gas and dust temperature increases from 10 to 500~K over 52 (fast), 203 (medium), or 1000 (slow) kyr according to the equation \textit{T(t) = 10 + $\kappa$t$^2$} as based upon the methods in \citet{Viti2004}, \citet{Garrod2006}, and \citet{Garrod2008}.  The values of \textit{$\kappa$} used in this work to warm from 10-500~K in the prescribed times for the fast, medium, and slow warm-ups are 1.96 $\times$ 10$^{-22}$, 1.2 $\times$ 10$^{-23}$, and 4.9 $\times$ 10$^{-25}$ K/s$^2$, respectively.

\begin{figure*}[!htb]
   \centering
      \includegraphics[width=\hsize]{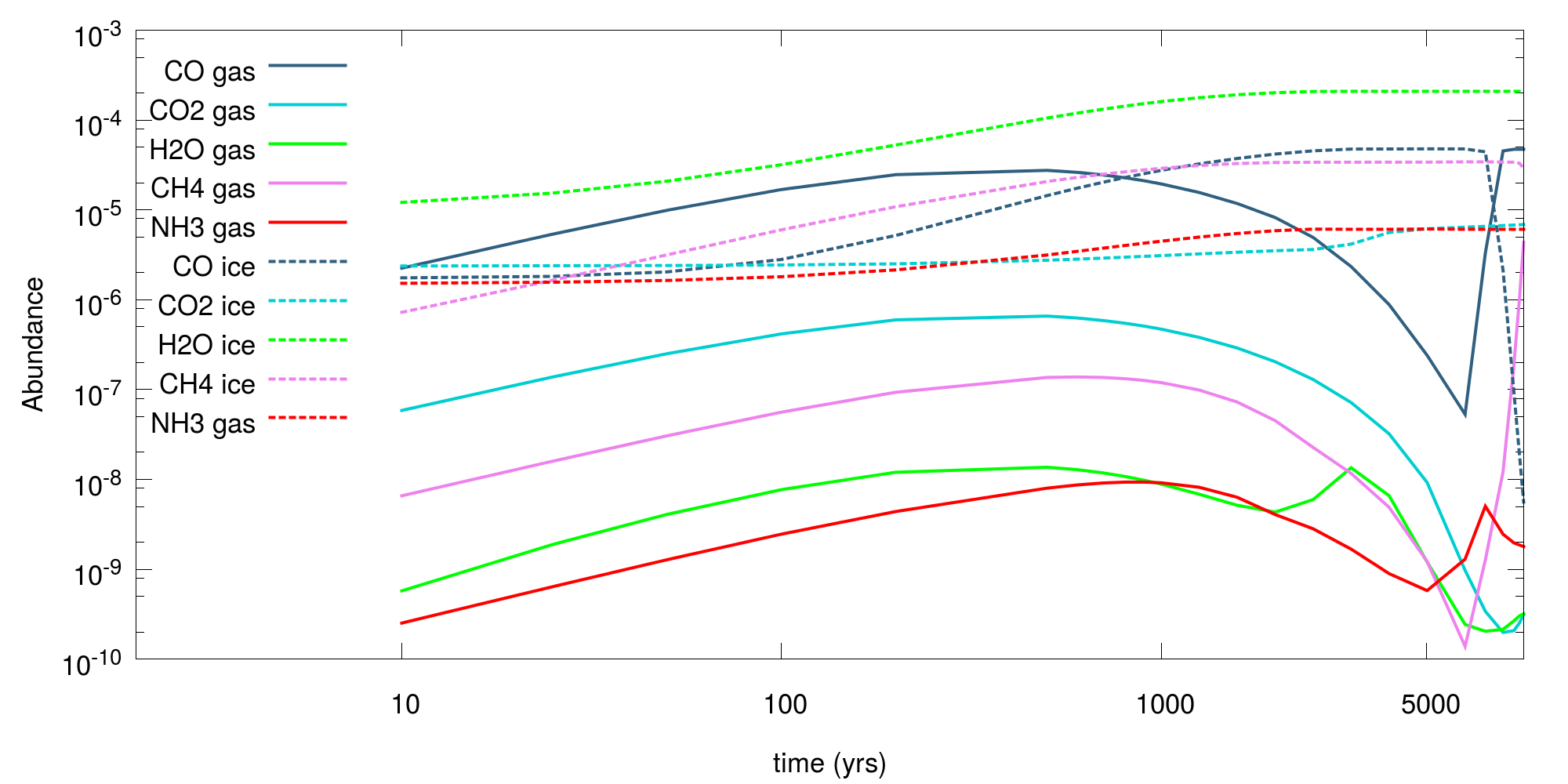}
      \caption{Abundances of simple species over time for a fast warm-up with a gas density of 10$^7$ cm$^{-3}$. Dashed lines show ice abundances and solid lines show gas abundances.}
         \label{icegas}
   \end{figure*}

\par The initial ice abundances for IC 2, 4, and 5 are from ice observations carried out by \citet{Gibb2004} of three high-mass star-forming regions, i.e., AFGL 2136, W33A, and NGC 7538 IRS9, respectively.  Initial conditions (IC) 1 and 3 are based on a lower limit of the water abundance of 10$^{-5}$ versus H$_2$ and an upper limit of water abundance of 10$^{-4}$ versus H$_2$ as suggested in \citet{Ewine2004}.  Other molecular abundances in IC 1 and 3 are then percentages of 5$\%$ (for NH$_3$, CH$_3$OH, and CH$_4$ ice) or 10$\%$ (for CO, CO$_2$, HCOOH, and H$_2$CO ice) of water.  The atomic gas abundances in our model are the result of subtracting the atoms that have gone into molecules from the typical gas abundances found in the pristine model input.  \textbf{Our approach differs from \citet{Garrod2008} in that their model includes a phase of initial collapse from diffuse cloud to dense core, thereby building up their ices in a model-dependent manner.} Our full gas and ice initial conditions are detailed in Appendix~\ref{appIC}. While we did not include molecular gas in our initial abundances, the chemistry quickly converts the free atoms into stable molecules (Figure~\ref{icegas}).  These gas-phase abundances are within an order of magnitude of reported abundances in starless and prestellar cores \citep{Prestellar2011,Evgenia2016,Vastel2016}.

\begin{figure*}[!htb]
   \centering
      \includegraphics[width=\hsize]{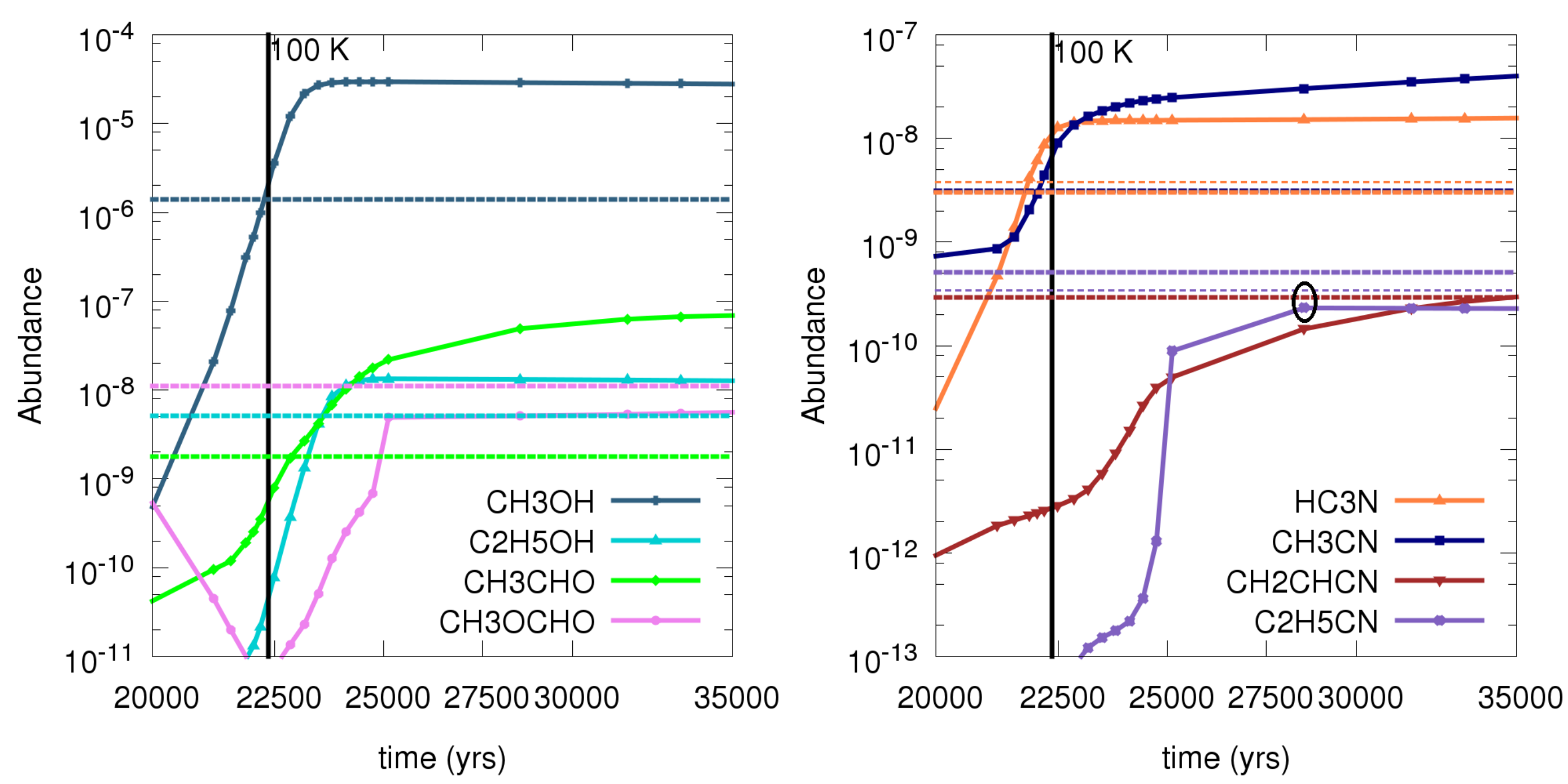}
      \caption{Fractional abundance for B3 fast warm-up (from 10-500~K) model at a density of 10$^7$ cm$^{-3}$ for the time period 20000-35000 yr with the fiducial model.  Oxygen-bearing species are shown to the left and nitrogen bearing to the right.  The y-axes have different scales.  All species are shown in the key with color-coded dashed horizontal lines showing the observed abundances for B3.  The thinner dashed lines indicate the upper limit for HC$_3$N and the lower limit for C$_2$H$_5$CN, as they are the species that best constrain the time span.  The black ellipse highlights the difference between the modeled abundance of C$_2$H$_5$CN and the lower limit of the observed abundance.  At the higher densities we tested, the model abundance of C$_2$H$_5$CN is lower.}
         \label{B3Fastlow}
   \end{figure*}

\subsection{Modeling approach}
First, we tested the fiducial model (\S~\ref{basic}): three different densities (based on the expected densities of our observed sources) at three different warm-up speeds based on expected gas warming speed around low-, intermediate-, and high-mass stars (discussed in \S~\ref{MassWarmup}) for the five initial conditions outlined above (\S~\ref{ICsec}).  After the basic test, we tested models that each changed one feature of the fiducial model for a fast warm-up (as expected for a high-mass source).  These were, excluding reaction-diffusion competition(\S~\ref{RDC}), changing the initial temperature to 15 or 25~K (\S~\ref{Ti}), running time at a high temperature (300~K) after the warm-up period (\S~\ref{Tf}), adding HCN to the initial ice (\S~\ref{HCN}), and raising the cosmic-ray ionization rate to 1$\times$10$^{-16}$ s$^{-1}$ and 6$\times$10$^{-16}$ s$^{-1}$ (\S~\ref{CRi}).  Modeled abundances for each test were compared to the observed abundances from \citet{Allen} shown in Table~\ref{abundancesTable}) and their associated upper and lower limits to constrain the time period during which all abundances could be reproduced by the model.

  \begin{table}[!bht]
  \centering
  \caption{Abundances vs. H$_2$ across G35.20 from \citet{Allen}.}
  \label{abundancesTable}
  {\small
  \begin{tabular}{cccccc}

  \hline\hline
  Species & Source A & Source B1 & Source B2 & Source B3 \\
  \hline
  CH$_3$OH & 5.0$\times$10$^{-7}$ & 6.2$\times$10$^{-7}$ & 6.7$\times$10$^{-7}$ & 1.4$\times$10$^{-6}$  \\
  C$_2$H$_5$OH & 3.0$\times$10$^{-9}$ & 9.4$\times$10$^{-10}$ & 3.1$\times$10$^{-10}$ & 5.2$\times$10$^{-9}$  \\
  CH$_3$CHO & 1.1$\times$10$^{-9}$ & 1.9$\times$10$^{-9}$ & 6.9$\times$10$^{-10}$ & 1.7$\times$10$^{-9}$  \\
  CH$_3$OCHO & 3.4$\times$10$^{-9}$ & 7.0$\times$10$^{-9}$ & 6.1$\times$10$^{-9}$ & 1.1$\times$10$^{-8}$  \\
  CH$_3$CN & 3.0$\times$10$^{-9}$ & 1.7$\times$10$^{-9}$ & 3.9$\times$10$^{-10}$ & 3.1$\times$10$^{-9}$  \\
  CH$_2$CHCN & 5.3$\times$10$^{-10}$ & $<$1$\times$10$^{-13}$ & $<$2$\times$10$^{-13}$ & 3.0$\times$10$^{-10}$  \\
  C$_2$H$_5$CN & 6.4$\times$10$^{-10}$ & $<$7$\times$10$^{-14}$ & $<$1$\times$10$^{-13}$ & 5.2$\times$10$^{-10}$ \\
  HC$_3$N & 5.1$\times$10$^{-10}$ & 2.4$\times$10$^{-10}$ & 5.9$\times$10$^{-11}$ & 3.1$\times$10$^{-9}$ \\
  \hline 
  \end{tabular}
  }
  \end{table}

%  NH$_2$CHO & 3.2$\times$10$^{-10}$ & 8.9$\times$10$^{-10}$ & 7.0$\times$10$^{-11}$ & 6.5$\times$10$^{-10}$  \\

\begin{figure*}[!htb]
   \centering
      \includegraphics[width=\hsize]{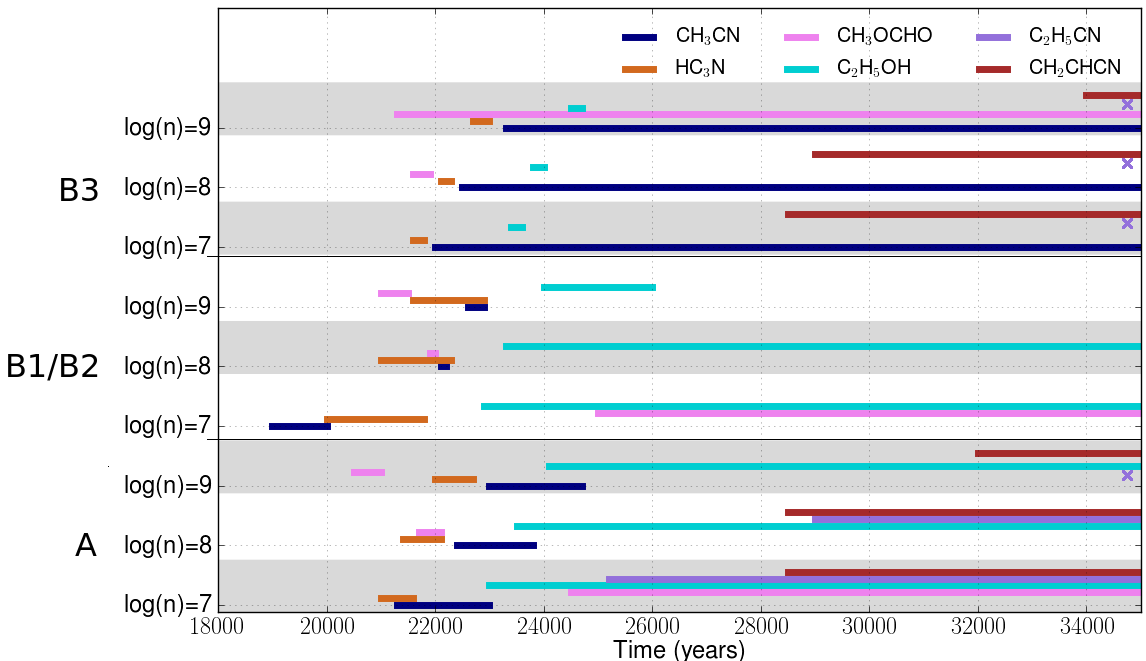}
      \caption{Time periods for which the observed abundances of HC$_3$N, CH$_3$CN, CH$_2$CHCN, C$_2$H$_5$CN, CH$_3$OCHO, and CH$_2$H$_5$OH are reproduced. The purple 'X' marks indicate that the abundance of C$_2$H$_5$CN is not reproduced for this source and gas density.}
         \label{barplot}
   \end{figure*}

   \begin{figure*}[!htb]
   \centering
      \includegraphics[width=\hsize]{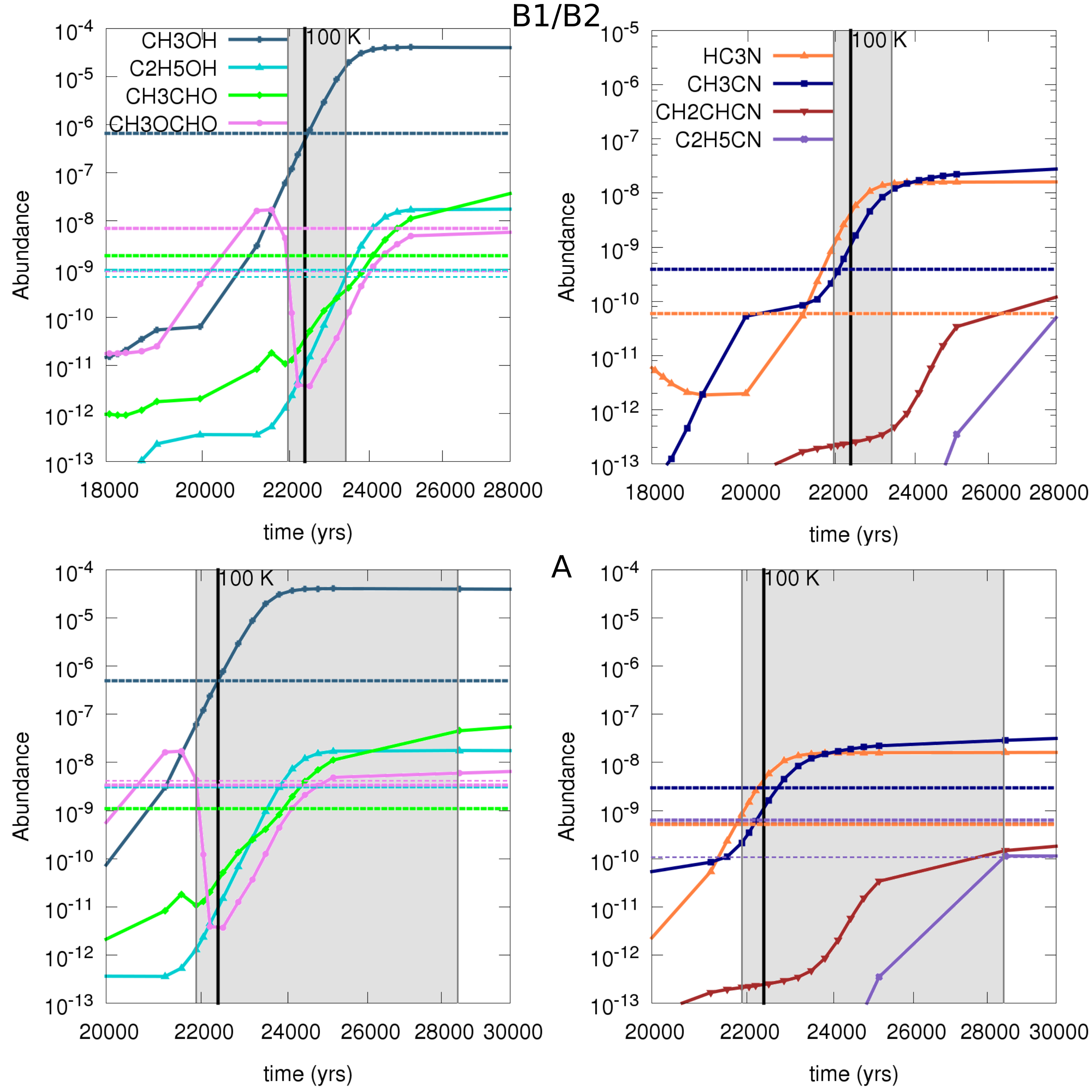}
      \caption{Best fit models of abundances vs. H$_2$ for CH$_3$OH, C$_2$H$_5$OH, CH$_3$CHO, CH$_3$OCHO, HC$_3$N, CH$_3$CN, CH$_2$CHCN, and C$_2$H$_5$CN using IC 5. G35.20 B1/B2 (top) is best fit by the fast model with a gas density of 10$^{8}$~cm$^{-3}$ over a time period of 1.3 kyr.  G35.20 A (bottom) is also best fit by the fast model with a gas density of 10$^{8}$~cm$^{-3}$ over a time period of 6.4 kyr.  The line colors for A are the same as B1/B2 as shown in the key.  All species are shown in the key with color-coded dashed horizontal lines showing the observed abundances for the source.  The thinner dashed lines indicate the upper limit and lower limit species that best constrain the time span.  The time ranges in which all abundances can be reproduced within the errors reported in Appendix~\ref{3520errors} are shaded.  We truncate the x-axis scale to better highlight the chemistry changes over the temperature range at which the COMs are released from the ice mantles.}
         \label{timefits}
   \end{figure*}

\begin{table*}[!htb]
\centering
\caption{Time ranges (in kyr) that fit observed abundances using the fiducial model in the lower abundance sources, B1/B2, those for the higher abundance source, B3, and the other hot core in this group, A.  Corresponding temperatures are also shown (in K).  Full details in Appendix~\ref{appIC}.}
\label{TimeRangeTable}
\begin{tabular}{cccccccc}
  \hline\hline
 &  & \multicolumn{2}{c}{\textbf{A}} & \multicolumn{2}{c}{\textbf{B1/B2}} & \multicolumn{2}{c}{\textbf{B3}} \\
Density & Warm-up time & Time range & Temperature & Time range & Temperature & Time range & Temperature \\
(cm$^{-3}$) & (kyr) & (kyr) & (K) & (kyr) & (K) & (kyr) & (K) \\
\hline
10$^{7}$ & 52 & 21.6-28.5 & 93-158 & 20.0-25.0 & 81-123 & \multicolumn{2}{c}{C$_2$H$_5$CN 2$\times$ too low} \\
10$^{8}$ & 52 & 22.1-28.5 & 97-158 & 22.0-23.3 & 96-107 & \multicolumn{2}{c}{C$_2$H$_5$CN 3$\times$ too low} \\
10$^{9}$ & 52 & \multicolumn{2}{c}{C$_2$H$_5$CN 5$\times$ too low} & 21.5-24.0 & 92-114 & \multicolumn{2}{c}{C$_2$H$_5$CN 10$\times$ too low} \\
10$^{7}$ & 203 & 85.0-97.3 & 94-121 & 75.5-90.0 & 76-105 & 84.5-97.5 & 93-116 \\
10$^{8}$ & 203 & 86.0-105.0 & 96-139 & 85.5-91.0 & 95-107 & 87.0-115.0 & 98-165 \\
10$^{9}$ & 203 & 88.7-103.0 & 102-134 & 88.0-94.5 & 100-114 & 89.6-103.0 & 104-134 \\
10$^{7}$ & 1000 & 365-472 & 75-119 & 365-425 & 75-98 & 375-475 & 78-120 \\
10$^{8}$ & 1000 & 415-490 & 94-127 & 410-455 & 92-111 & 420-490 & 96-127 \\
10$^{9}$ & 1000 & 430-500 & 100-132 & 420-460 & 96-113 & 435-500 & 102-132 \\
\hline
\end{tabular}
\end{table*}

\section{Results}
We aim to constrain the time periods during which the models reasonably reproduce the observed abundances within observed error limits (detailed in Table~\ref{abundancesTable}) of the following molecules: cyanides (CH$_3$CN, CH$_2$CHCN, C$_2$H$_5$CN), cyanoacetylene (HC$_3$N), methanol (CH$_3$OH), methyl formate (CH$_3$OCHO), acetaldehyde (CH$_3$CHO), and ethanol (C$_2$H$_5$OH).  For a time period to be an acceptable fit, its duration must be less than half a disk rotation period ($<$5~kyr). The focus species include all of the cyanides observed in G35.20 and most of the complex organic oxygen-bearing species.  The abundances in Table~\ref{abundancesTable} were determined using detailed modeling of ALMA observations of spectral lines from the four continuum points in Figure~\ref{ContPts3520} \citep{Allen} using the software XCLASS \citep{XCLASS}, assuming local thermal equilibrium (LTE).  The key result from this analysis was that continuum peak B3 showed higher abundances of almost all modeled species, but especially of those containing the CN group.  Two in particular, vinyl and ethyl cyanide (CH$_2$CHCN, C$_2$H$_5$CN), were not detected at continuum peaks B1 and B2, giving an upper limit to their abundances of 1$\times$10$^{-13}$ with respect to H$_2$. Upper and lower limits for the XCLASS modeling results can be found in Appendix~\ref{3520errors}.
\par We find little variation among the different starting conditions (see Figure~\ref{ICs}), so in the following analysis we use the initial ice composition of IC5 as NGC 7538 IRS9 has a similar bolometric luminosity and distance to G35.20 (4 $\times$ 10$^4$ L$_{\sun}$ and 2.7 kpc for NGC 7538 IRS9 versus 3 $\times$ 10$^4$ L$_{\sun}$ and 2.2 kpc for G35.20).

\begin{figure*}[!htb]
   \centering
      \includegraphics[width=\hsize]{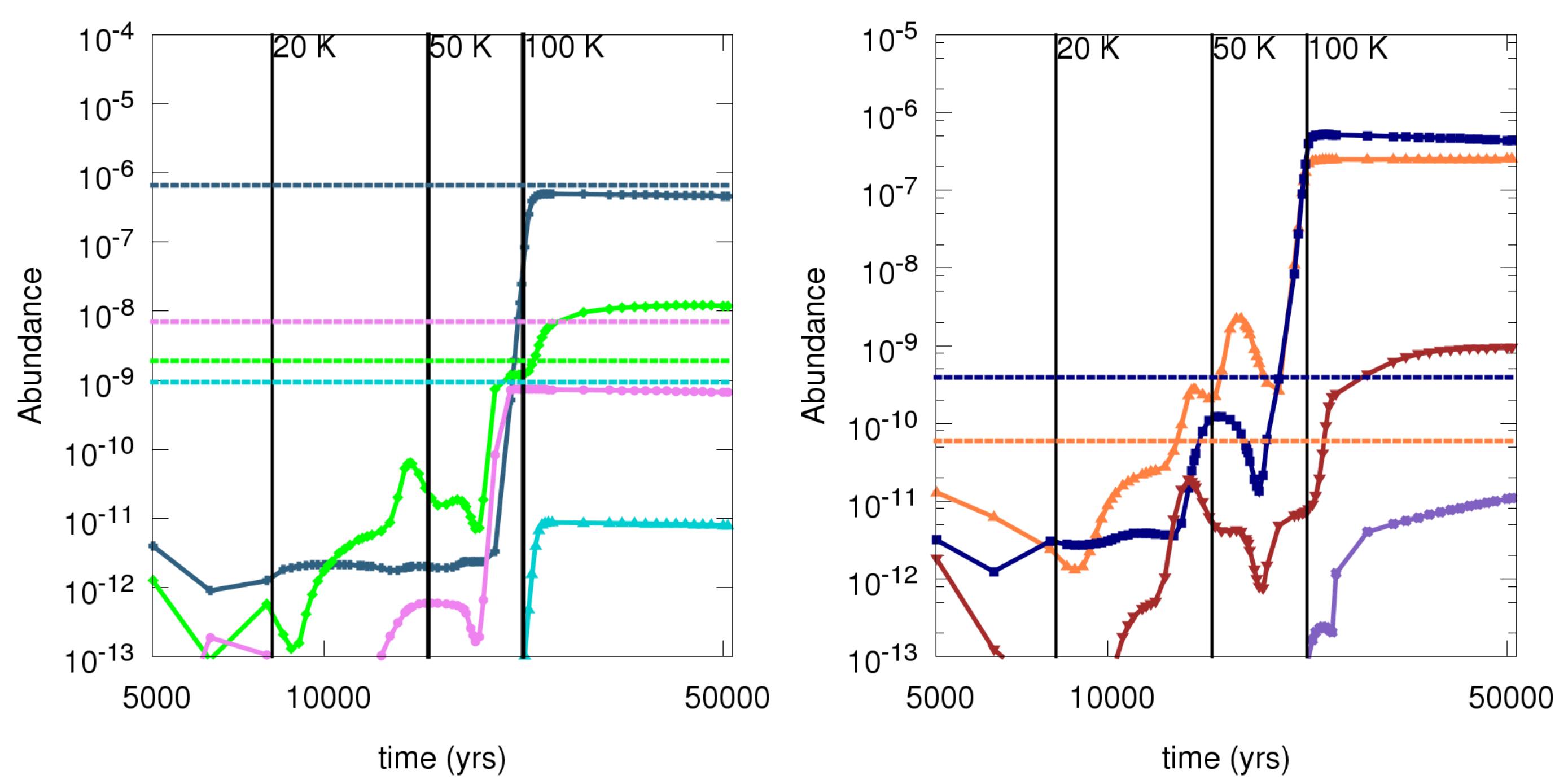}
      \caption{Fractional abundance for B1-2 fast warm-up (from 10-500~K) model at a gas density of 10$^7$ cm$^{-3}$ for the time period 5-50 kyr without reaction-diffusion competition.  All species are color coded as in Figure~\ref{timefits} with horizontal lines showing the observed abundances for B1/B2.  The abundances of CH$_3$OH, CH$_3$OCHO, and C$_2$H$_5$OH are not reproduced.}
         \label{RDCoff}
   \end{figure*}

\subsection{Fiducial model}
\label{basic}
In our fiducial model, we begin with gas and dust at 10~K and the initial ice and gas conditions of IC5, then warm the gas and dust at fast, medium, and slow speeds (detailed in \S~\ref{chemModel}) to 500~K at a constant gas density. All of the tests in the following subsections begin with the conditions of this fiducial model varying one parameter.  Time ranges and corresponding gas temperature ranges for all fits are summarized in Table~\ref{TimeRangeTable}. 
\par Fast warm-up models can reproduce all of the abundances observed for peaks B1/B2 and the C$_2$H$_5$CN abundance can be reproduced in core A for densities of n = 10$^{7}$ and 10$^{8}$ cm$^{-3}$. The model C$_2$H$_5$CN abundance at peak B3 cannot be reproduced by the fiducial model, although at a density of 10$^{7}$~cm$^{-3}$ it is 1.4$\times10^{-10}$ lower (50$\%$) than the minimum observed abundance (see Figure~\ref{B3Fastlow}). This difference is significantly larger than the tolerance for the model (10$^{-13}$) and is therefore not a fit. A summary of the time periods where the observed abundances of HC$_3$N, CH$_3$CN, CH$_2$CHCN, C$_2$H$_5$CN, CH$_3$OCHO, and CH$_2$H$_5$OH are reproduced in a fast warm-up for all sources and gas densities is shown in Figure~\ref{barplot}.
\par Medium-speed warm-up models can reproduce the C$_2$H$_5$CN abundance observed in source B3 at late times (after 97~kyr).  The time period required to reproduce all observed abundances is $>$13~kyr, which is longer than a disk rotation period. Abundances in source B1/B2 can be reproduced in a medium warm-up in $\sim$5.5~kyr. See Appendix~\ref{timeperiods} for tables detailing the times at which observed abundances are replicated by the fiducial model and further plots of abundance over time.
\par Slow warm-up models for all three densities can reproduce all of the observed abundances.  The time periods needed to reproduce the observed abundances are shorter for n = 10$^{8}$ and 10$^{9}$~cm$^{-3}$, although still very long ($>$40~kyr).  For peaks B1/B2 the shortest time range is 40~kyr at n = 10$^{9}$~cm$^{-3}$ (temperature range 96-113~K). The shortest time range for peak B3 is 65~kyr, corresponding to a temperature range of 102-132~K at n = 10$^{9}$~cm$^{-3}$. These time ranges are not reasonable, as the gas in the disk would have made several revolutions during such a long period.
\par The fiducial model fits peaks B1/B2 and core A very well using a fast warm-up.  Abundances toward peaks B1/B2 can even be reproduced within a time period of 1.3 kyr at a gas density of 10$^{8}$~cm$^{-3}$.  Core A requires a longer time period of 6.4 kyr, but is still well fit at a gas density of 10$^{8}$~cm$^{-3}$ with a fast warm-up.  The best fit models for B1/B2 and A are shown in Figure~\ref{timefits}.  The abundances of C$_2$H$_5$CN toward peak B3 cannot be reproduced using a fast warm-up, but the shortest time period (13 kyr) that reproduces all abundances is using a medium warm-up at a gas density of 10$^{7}$~cm$^{-3}$.  This is too long kinematically (the disk rotational period is $\sim$10 kyr), and we expect it to be a high-mass source with a fast warm-up time because it has a high luminosity, cluster of OH masers \citep{OHmasers1999}, and a high kinetic temperature ($\sim$300~K) together with a high deuterium fraction implying that it has recently heated up very quickly \citep{Allen}. As our model does not use any reactions with deuterium, we can only assume that the model abundances may differ from those listed if these reactions were included.

\subsubsection{Constraint species}
For nearly all warm-up speeds and densities, the lower time range is constrained by the HC$_3$N abundance in core A and source B3 and by CH$_3$CN in B1/B2.  Where this is not the case, CH$_3$OCHO is the lower abundance constraint. The upper time range is constrained by the C$_2$H$_5$CN abundance for source B3 and core A in medium and slow warm-ups and by CH$_2$CHCN in fast warm-ups, where source B3 cannot be reproduced because of the high C$_2$H$_5$CN abundance.  The C$_2$H$_5$OH abundance provides the upper time range constraint for  sources B1/B2 in most cases, but the CH$_3$OCHO abundance provides the upper limit for the fast warm-up at 10$^{7}$~cm$^{-3}$ and the slow warm-up at 10$^{8}$~cm$^{-3}$ and CH$_3$CHO is the upper constraining species for the medium and slow warm-ups at 10$^{9}$~cm$^{-3}$.

\par When investigating the time ranges for B1/B2 observations for the abundances of the unobserved cyanides (CH$_2$CHCN and C$_2$H$_5$CN), we see that they are very low (between 10$^{-10}$ and 10$^{-14}$ for all models).  The abundances of these species increase rapidly in a short space of time.  The most dramatic is C$_2$H$_5$CN which jumps up 2 orders of magnitude within 1000 years in the fast warm-up at n = 10$^{7}$ cm$^{-3}$ (a temperature change of $\sim$10~K).

\subsection{Reaction-diffusion competition excluded}
\label{RDC}
Reaction-diffusion competition is a mechanism used in chemical modeling to allow grain surface reactions with energy barriers to proceed more easily.  This mechanism compares the relative timescales between the reaction of two species and their diffusion to determine which process will occur \citep{Cuppen}.  Because reaction-diffusion competition may be overexpressed in a two-phase chemical model (gas and ice), we modeled a test case without it.  In this case, key species such as CH$_3$OH, CH$_3$OCHO, and C$_2$H$_5$OH are underproduced by as much as 2 orders of magnitude compared to the lower limit abundances for any of our observed sources.  Figure~\ref{RDCoff} demonstrates one instance where all three of these species are underproduced for B1-2.  More efficient grain-surface chemistry facilitated by reaction-diffusion competition is required to better match the observations, which should be tested in more detailed three-phase chemical models.

   \begin{figure}[!htb]
   \centering
      \includegraphics[height=0.85\vsize]{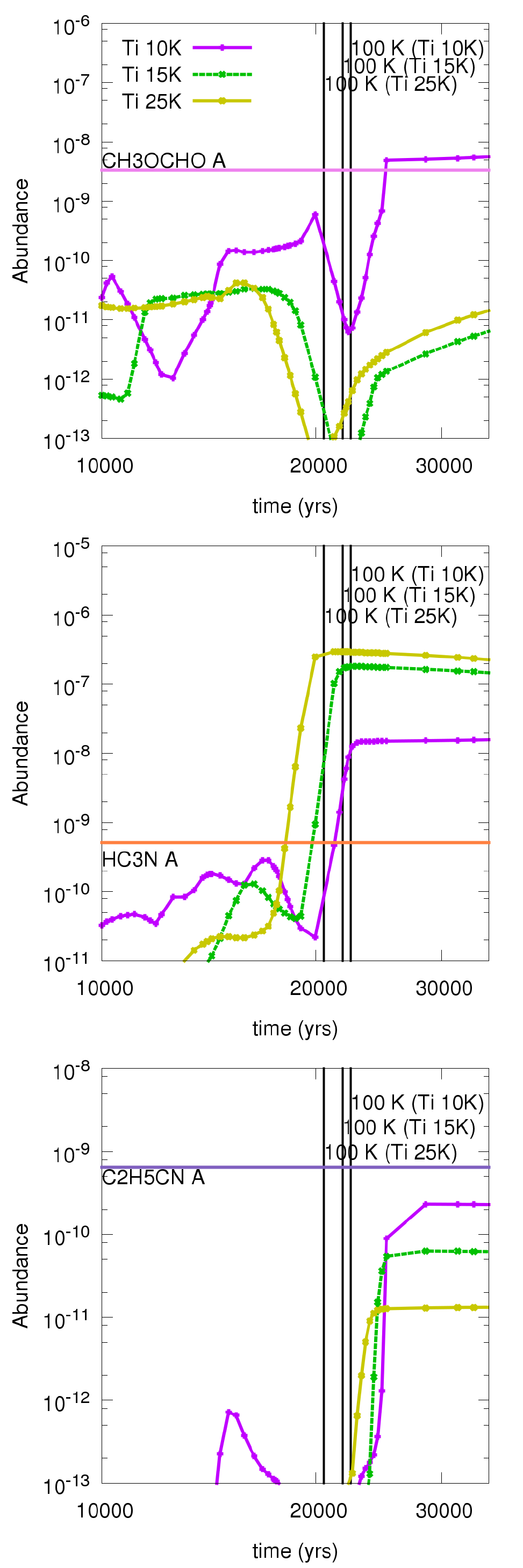}
      \caption{Comparing the fractional abundances for three constraining species, CH$_3$OCHO (top), HC$_3$N (middle), and C$_2$H$_5$CN (bottom), using different initial temperatures (10~K in magenta, 15~K in green, and 25~K in yellow) at a gas density of 10$^{7}$~cm$^{-3}$ for a fast warm-up. The vertical lines show the time corresponding to a temperature of 100~K for each initial temperature.  The observed abundances (horizontal lines) for G35.20 A are shown for reference.}
         \label{compareT}
   \end{figure}
\clearpage
\subsection{Varying the initial temperature}
\label{Ti}
It is plausible that for high-mass stars forming in a cluster, possibly sequentially, an initial temperature of 10 K is an underestimation \citep{Tieftrunk1998}.  For this reason we also modeled the chemistry of dense gas warming up from 15 and 25~K.  Looking at the changes in abundance for constraining species, CH$_3$OCHO, HC$_3$N, and C$_2$H$_5$CN, we see that increasing starting temperatures decrease the abundances of CH$_3$OCHO and C$_2$H$_5$CN but increase the abundance of HC$_3$N (see Figure~\ref{compareT}).  For C$_2$H$_5$CN, longer times at as a low temperature allows more to form in the ice, to be later released into the gas phase.
\par Because the warm-up is exponential, starting at 15~K rather than 10~K results in 6000 years less at a low temperature (and 10000 for 25~K) for the fast warm-up.  Since the formation path to C$_2$H$_5$CN is mainly in the ice it appears that time at a low temperature is critical.  This is demonstrated as well in the medium and slow warm-ups in which high abundances of C$_2$H$_5$CN are made as these models spend a very long time at low temperatures.  The temperature range between 15 and 30~K is critical for grain-surface reactions because the dust temperature determines the sticking efficiency of volatile species (such as H, H$_2$, and CO).  At higher temperatures hydrogenation pathways (such as those that lead to C$_2$H$_5$CN) are less likely to occur. 

   \begin{figure}[!htb]
   \centering
      \includegraphics[height=0.85\vsize]{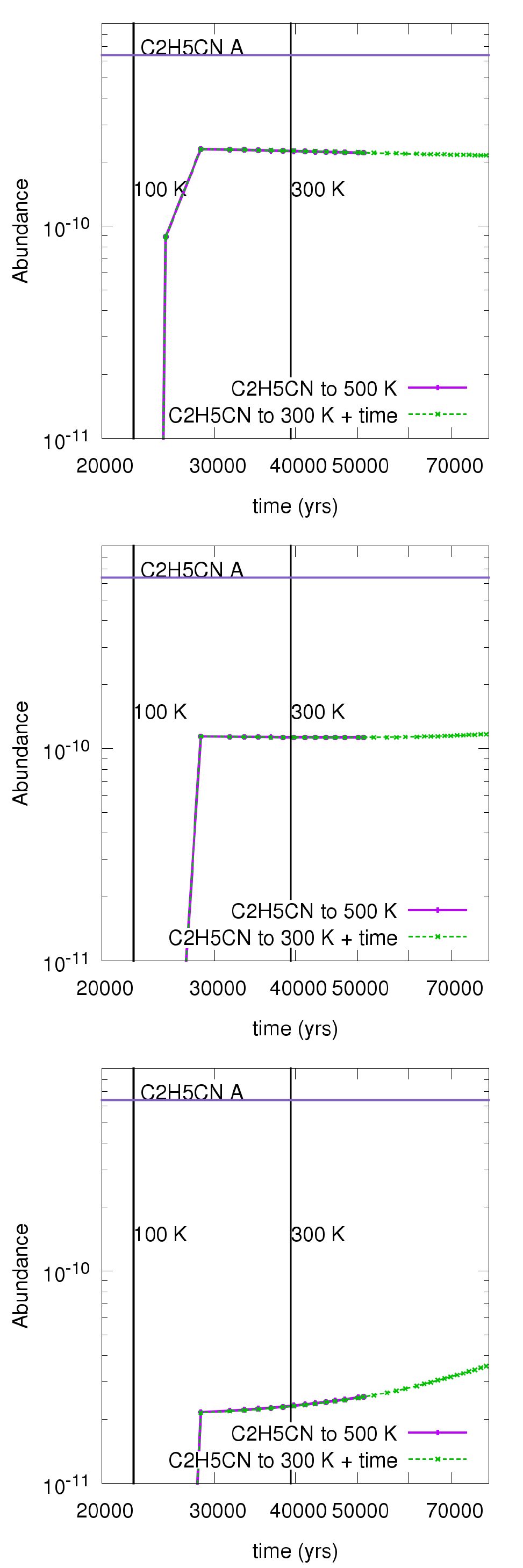}
      \caption{Fractional abundances for C$_2$H$_5$CN comparing warming up to 500~K with warming up to 300~K then continuing at a constant temperature at densities of 10$^{7}$~cm$^{-3}$ (top), 10$^{8}$~cm$^{-3}$ (middle), and 10$^{9}$~cm$^{-3}$ (bottom) for a fast warm-up. The observed abundance (horizontal lines) for G35.20 A is shown for reference.}
         \label{compare300}
   \end{figure}

   \begin{figure*}[htb]
   \centering
      \includegraphics[width=\hsize]{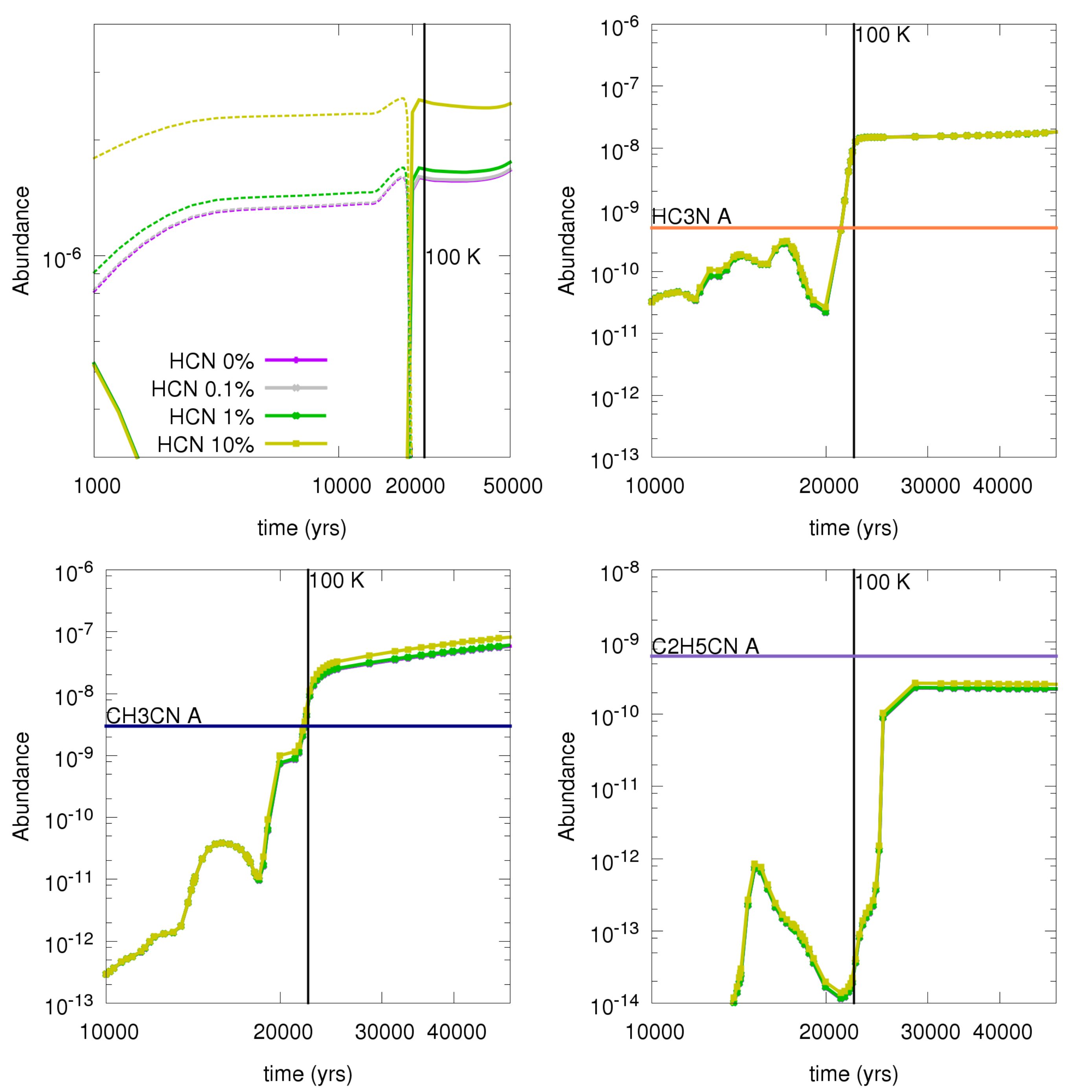}
      \caption{Fractional abundances for nitrogen-bearing species with a gas density of 10$^{7}$~cm$^{-3}$ and a fast warm-up.  HCN gas (solid lines) and ice (dashed lines) abundances with the color key for all four panels (top left), HC$_3$N gas (top right), CH$_3$CN gas (bottom left), and C$_2$H$_5$CN gas (bottom right) abundances over time are shown for four initial HCN ice abundances (0, 0.1, 1, and 10$\%$).  The observed abundances (horizontal lines) for G35.20 A are shown for reference.}
         \label{compareHCN}
   \end{figure*}

\subsection{Continuing with constant high temperature gas-phase chemistry}
\label{Tf}
To investigate the effect of high temperature gas-phase chemistry on our final abundances in the fast warm-up, we modeled warming up the dense gas to 300~K, then continued at that temperature for an additional 40~kyr.  As C$_2$H$_5$CN is the only species that cannot be fit for source B3, we focus on the abundance of this species produced at different densities with extra time to perform gas-phase chemistry.  In Figure~\ref{compare300}, we see that the abundances produced after 300~K do not deviate from those when the gas continues to warm to 500~K.  In the extra time, abundances only increase at the highest gas density and then by 36$\%$ ($\sim 1\times10^{-11}$), which does not reproduce the observed minimum abundance.

\subsection{HCN as an initial ice species}
\label{HCN}
HCN has been observed in cometary ice \citep{{CometIce1},{CometIce2}} and is expected to occur in ices around protostars but has not yet been detected \citep{Boogert2015}.  To test the effect of including HCN in the ice, we modeled the following three additional initial abundances of HCN ice: 0.1$\%$, 1$\%$, and 10$\%$ versus H$_2$O.  An abundance of 0.1$\%$ reflects the observed abundance in cometary ice (0.08-0.5$\%$), but the higher abundances were used to test if there was any increase in our CN-bearing species using an unrealistic concentration of HCN.  In Figure~\ref{compareHCN} we see that the constraining species are barely affected by this change in HCN abundance, while the HCN gas abundances are directly affected.  We conclude that HCN is not an important progenitor to any of the nitrogen-bearing species that we are focusing on and for the range of physical conditions explored in this work.

   \begin{figure*}[htb]
   \centering
      \includegraphics[width=\hsize]{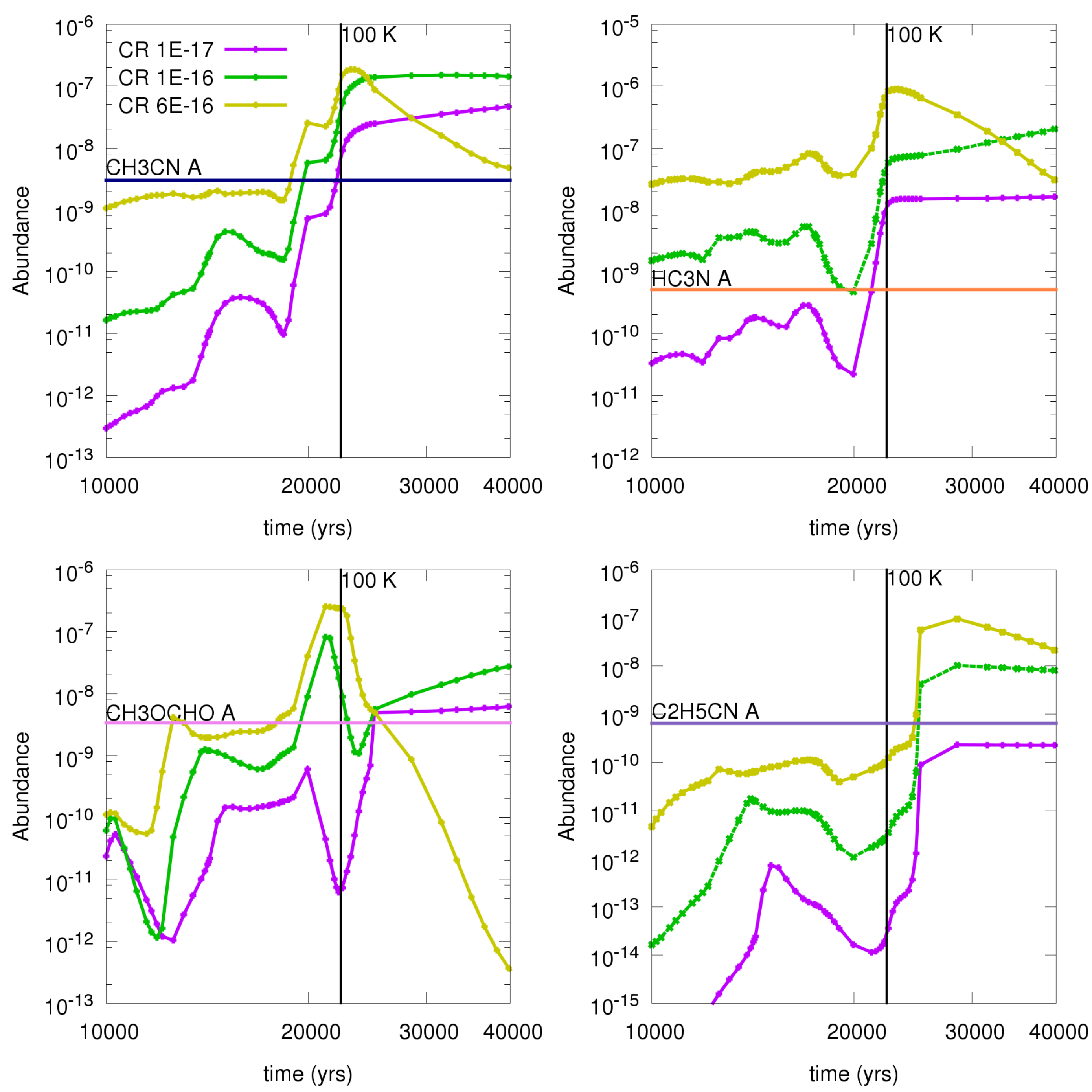}
      \caption{Comparing fractional abundances for four constraining species with different cosmic-ray ionization rates with a gas density of 10$^{7}$~cm$^{-3}$ for a fast warm-up.  CH$_3$CN with the color key for all four panels (top left), HC$_3$N (top right), CH$_3$OCHO (bottom left), and C$_2$H$_5$CN gas (bottom right) abundances over time are shown for three cosmic-ray ionization rates (1.3$\times10^{-17}$, 1$\times10^{-16}$, and 6$\times10^{-16}$ s$^{-1}$). The observed abundances (horizontal lines) for G35.20 A are shown for reference.}
         \label{compareCR}
   \end{figure*}

\begin{table}[h]
\centering
\caption{Time ranges (in kyr) that are needed fit observed abundances in the lower abundance sources, B1/B2, those for the higher abundance source, B3, at a cosmic-ray ionization rate of 1$\times10^{-16}$ s$^{-1}$ in a fast warm up.  Corresponding temperatures are also shown (in K).}
\label{CRresults}
\begin{tabular}{c|cc|cc}
  \hline\hline
 & \multicolumn{2}{c}{\textbf{B1/B2}} & \multicolumn{2}{c}{\textbf{B3}} \\
Density & Time range & T$_{gas}$ & Time range & T$_{gas}$ \\
(cm$^{-3}$) & (kyr) & (K) & (kyr) & (K) \\
\hline
10$^{7}$ & 19-22.75 & 74-101 & 21.25-25 & 90-123 \\
10$^{8}$ & 20-23 & 81-105 & 21.9-25.4 & 96-127 \\
10$^{9}$ & 22.3-23.6 & 99-110 & 22.7-26 & 108-132 \\
\hline
\end{tabular}
\end{table}

\subsection{Varying the cosmic-ray ionization rate}
\label{CRi}
The fiducial model uses a low cosmic-ray ionization rate commonly used in chemical modeling of 1.3$\times10^{-17}$ s$^{-1}$, which is low compared with more distant observed star-forming regions \citep{Indriolo2015}.  We modeled the chemistry over time for two higher cosmic-ray ionization rates, 1$\times10^{-16}$ s$^{-1}$ and 6$\times10^{-16}$ s$^{-1}$ , to be comparable to the mean and uppermost values from \citet{Indriolo2015}.  When comparing the changes in abundance for constraining species, CH$_3$CN, CH$_3$OCHO, HC$_3$N, and C$_2$H$_5$CN, we see that a higher cosmic-ray ionization rate increases their abundances, although after $\sim$ 25~kyr the abundances for a cosmic-ray ionization rate of 6$\times10^{-16}$ s$^{-1}$ drop sharply (see Figure~\ref{compareCR}).  This sharp drop in our key species is due to either a high abundance of H$_3$O$^{+}$ and HNCH$^{+}$ in the case of CH$_3$CN, CH$_3$OCHO, and HC$_3$N, or dissociation by cosmic rays for C$_2$H$_5$CN. A cosmic-ray ionization rate of 1$\times10^{-16}$ s$^{-1}$ presents a solution that fits the observed abundances in source B3 with a fast warm-up.  
\par
\begin{figure*}[htb]
   \centering
      \includegraphics[width=\hsize]{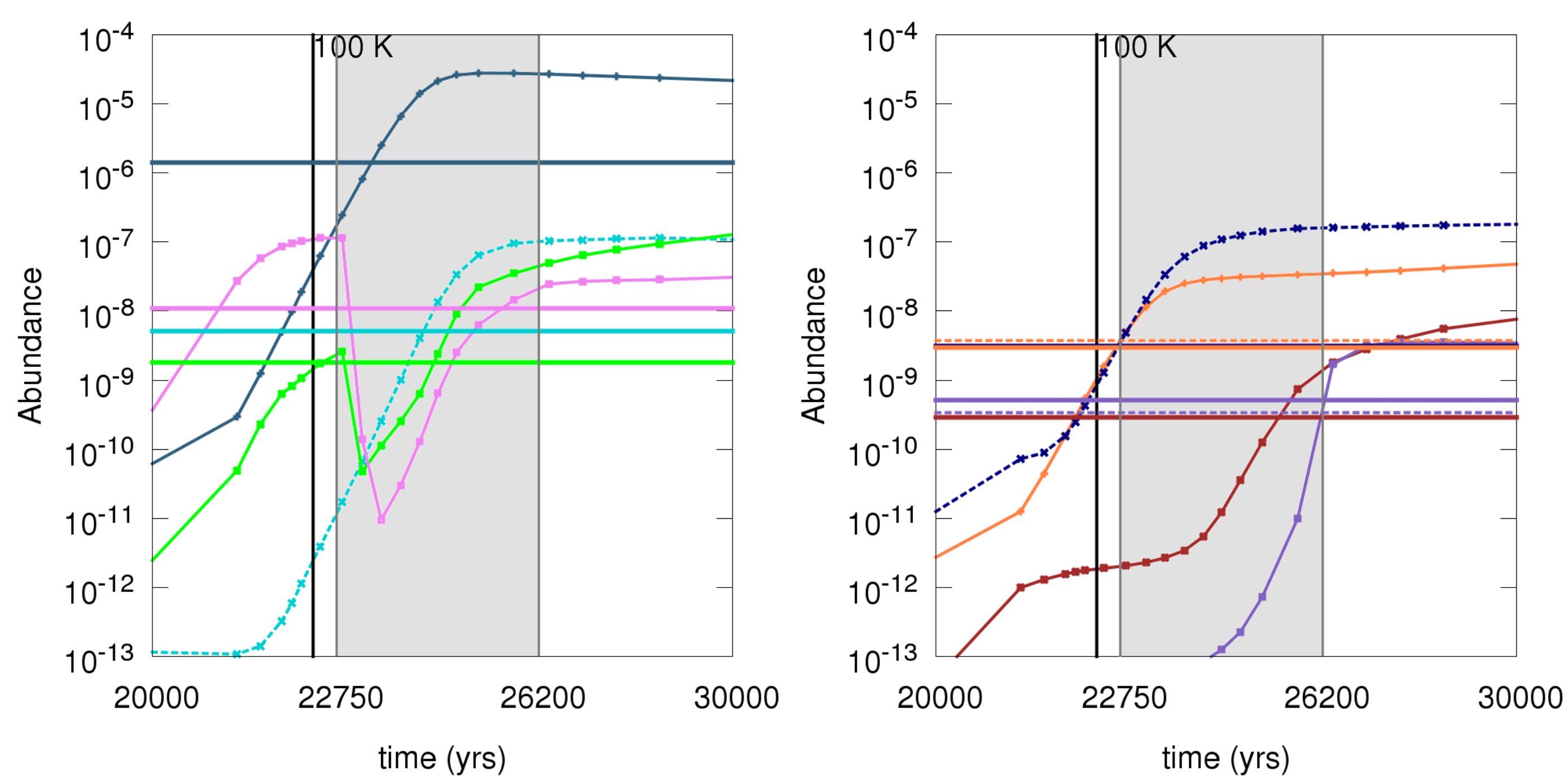}
      \caption{Fractional abundances for source B3 with a cosmic-ray ionization rate of 1$\times10^{-16}$ s$^{-1}$, a gas density of 10$^{9}$~cm$^{-3}$ for a fast warm-up. All species are shown in the key with color-coded dashed horizontal lines showing the observed abundances for source B3.  The thinner dashed lines indicate the upper limit for HC$_3$N and the lower limit for C$_2$H$_5$CN, as they are the species that constrain the time span.  The best fit time period of 3.4 kyr is shaded.  The colors are coded as in Figure~\ref{timefits}.}
         \label{B3CR}
   \end{figure*}

 The shortest time period that fits the observed abundances in source B3 is 3.3 kyr in a fast warm-up with a cosmic-ray ionization rate of 1$\times10^{-16}$ s$^{-1}$ and a gas density of 10$^{9}$~cm$^{-3}$ (Figure~\ref{B3CR}).  The observed abundances of A and B1-2 are also well fit with a cosmic-ray ionization rate of 1$\times10^{-16}$ s$^{-1}$.  A rate of 6$\times10^{-16}$ s$^{-1}$ raises the modeled abundance of HC$_3$N such that it no longer fits any of the observed abundances and so is not a viable solution for the model assumptions and parameters explored here.  Table~\ref{CRresults} summarizes the time ranges where the model abundances fit the observed abundances at a cosmic-ray ionization rate of 1$\times10^{-16}$ s$^{-1}$, with corresponding temperatures.  It is clear that for the same H$_2$ density and cosmic-ray ionization rate, there is a small time overlap between sources B1/B2 and source B3, and B3 is always a few thousand years older than B1/B2.

\subsection{Dominant formation routes}
We studied the reactions behind each of our eight focus species to determine whether they were formed mostly through ice processing and sublimation, through gas-phase formation following the sublimation of their precursors, or a mixture of both.

\par CH$_3$OH, HC$_3$N, and C$_2$H$_5$CN are predominantly produced on the grain surfaces then sublimated with little to no gas-phase production.  Significant amounts of CH$_3$CN are produced on the grain surface, but after sublimation gas-phase processes increase the abundance of CH$_3$CN gas to $\sim$8~times the maximum ice abundance. C$_2$H$_5$OH is also produced predominantly in the ice, but gas processes double the maximum ice abundance.

\par The ice and gas-phase abundances of CH$_3$CHO are unusual in that the ice abundance drops sharply around 63-70~K in the fast warm-up (at different densities).  This appears to coincide with an increase in CH$_3$OH and CH$_3$OCHO gas abundances. At this temperature in the model, grain surface CH$_3$CHO reacts with CH$_2$OH to form either C$_2$H$_5$OH and HCO or CH$_3$ and CH$_2$OHCHO in the ice, or it reacts with NH$_2$ to form NH$_3$ and CH$_3$CO in the ice as well.  CH$_2$OHCHO is important in forming CH$_3$OH and CH$_3$OCHO on grain surfaces. The CH$_3$CHO gas abundance does not increase until the temperature reaches $\sim$100~K.  At that temperature, the main production pathway is through neutral-neutral reactions between CH$_3$OH and CH in the gas phase.  So despite the significant abundances of CH$_3$CHO that are produced in the ice, very little of this sublimates into the gas phase.  The CH$_3$CHO gas is mainly a product of CH$_3$OH and CH.

\par CH$_3$OCHO is made abundantly in the ice, but reacts with OH in the ice to form COOCH$_3$ and water ice by hydrogen abstraction. COOCH$_3$ is hydrogenated in the ice and the resulting CH$_3$OCHO is released to the gas (H $_{(ice)}$ + COOCH$_3$ $_{(ice)}$ = CH$_3$OCHO).  This is the main mechanism for creating CH$_3$OCHO in the gas rather than sublimation of CH$_3$OCHO from the ice or formation in the gas.

\par CH$_2$CHCN is another species that can be made at low fractional abundances (10$^{-12}$-10$^{-10}$) in the ice and gas.  Ice phase CH$_2$CHCN is dominantly produced through the dissociation of C$_2$H$_5$CN by cosmic rays, whereas the gas-phase formation route is the reaction of CN with either C$_2$H$_4$ or CH$_3$CHCH$_2$. After a temperature of $\sim$115~K, the CH$_2$CHCN in the ice is sublimated and adds to the gas-phase abundance.

\section{Discussion}

\subsection{General test differences}
The first interesting result is that our model abundances are almost independent of initial ice conditions that we used, which are constrained by the observations of \citet{Gibb2004} and theoretical abundance ratios from \citet{Ewine2004}.  The range of ice abundances in our different IC sets is not large, but it is based on real sources.  The models suggest that the initial ice composition is not crucial to modeling the chemical composition of a later state, thereby showing that the warm-up stage determines the composition of the hot core. It is possible that adding molecular gas to the initial conditions would have an effect on the final abundances; however, such a parameter-space exploration of the initial gas composition requires a dedicated suite of models and is beyond the scope of this work. This should be carried out in the future.

\par The second interesting result is a much debated topic in chemical modeling \citep{Cuppen}: the importance of approximating reaction-diffusion competition in rate equation based models.  In this work, in order to reproduce the lower limit abundances of the species that we focus on from G35.20, reaction-diffusion competition is needed, otherwise oxygen-bearing complex organic species are underproduced by as much as 2 orders of magnitude. Reaction-diffusion competition has also been shown to be necessary in the work of \citet{Ruaud2016} and \citet{Quenard2018} among others, but it is important to note that gas-phase reactions for complex organic molecules in chemical networks have also been shown to be incomplete \citep{Balucani2015}.

\par The modeled gas temperatures at which the abundances are reproduced are lower than the kinetic temperatures (300~K for source B3, 285~K for core A, 160~K for source B1, 120~K for source B2) determined in \citet{Allen}.  The gas temperatures from our chemical model are 110-130~K for source B3, 100-110~K for sources B1/B2, and 100-130~K for core A.  As we have demonstrated that increasing the temperature and running the chemistry for longer does not significantly affect the final abundances, then the reproduction of the observed abundances at lower temperatures is an advantage.

\subsection{Reproducing source B3}
In our fiducial model, the warm-up time is the most significant factor in reproducing the abundances of the observed species.  The high abundances of C$_2$H$_5$CN seen in source B3 cannot be produced in a fast warm-up in our fiducial model.  While relatively short time ranges can be found to reproduce the abundances seen in sources B1/B2 in any of our models (with the shortest time range of 1.3 kyr from a fast model at 10$^{8}$~cm$^{-3}$), those of source B3 can only be reproduced in medium or slow warm-up models.  These longer warm-up times imply a lower mass protostar.   Observational evidence points to source B3 being associated with a high-mass protostar: there are numerous masers about its position and the kinetic temperature is high ($\sim$300~K).  On the other hand, a high deuterium fraction (13$\%$) for CH$_3$CN \citep{Allen} indicates that it was recently very cold and therefore needs a faster warm-up time.  

\par We reproduced the observed abundances in source B3 well using a cosmic-ray ionization rate a few times higher than in the standard value \citep{ffsvdtcr} for the interstellar medium.  A higher mean cosmic-ray ionization rate of 1.78$\times$10$^{-16}$~s$^{-1}$ was found in observations by \citet{Indriolo2015} and our models seem to agree with this higher rate. Source B1/B2 is also well reproduced within a 1.3~kyr time period using a cosmic-ray ionization rate of 1$\times$10$^{-16}$~s$^{-1}$, without the abundances of the complex cyanides becoming too high.

\par It is possible that adding gas-phase reactions forming C$_2$H$_5$CN to the network will make it possible to reproduce the higher abundances seen in source B3 with the fiducial model, as the network currently contains no gas-phase reactions to produce this species.  Such reactions are not often tested in the laboratory as cyanides are dangerous to work with, but it would be extremely useful for labs to test these reactions in the future to improve the chemical networks. 

%\par The time periods in fast warm up models that reproduce the abundances seen in A and B1/B2 are small enough to be within the beam of the observations (---the beam is 0.4$''$ which is 880~AU with a rotation speed of 3.5 km/s this gives about 1200 years crossing time for the beam---).

\subsection{Warm-up times}
\label{MassWarmup}
The Garrod models \citep{Garrod2006,Garrod2008} and their predecessor models \citep{viti} derive their warm-up times from the work of \citet{BM96} (BM96 from here on).  In BM96 work, the contraction times for different masses of stars (from 0.8-60 M$_{\sun}$) are determined under the assumption that the accretion rate is between 10$^{-5}$ and 10$^{-4}$~M$_{\sun}$yr$^{-1}$. It has been reported more recently that mass accretion rates can be as high as 10$^{-3}$~M$_{\sun}$yr$^{-1}$ \citep{Tan2014}, although this may be episodic.  \citet{Hosokawa2009} found that these high accretion rates led to pre-main sequence stars with larger-than-typical radii.  \citet{Macla2017} has recently reported observational evidence for this in M17. In any case, our fast, medium, and slow warm-up times correspond to 60, 15, and 6~M$_{\sun}$ objects from the original BM96 paper, considered to be very high, high, and intermediate mass sources.  If we assume that the accretion rate of our objects is ten times higher and decrease the contraction times of the BM96 objects accordingly, that gives more reasonable stellar masses of 8, 4, and 1~M$_{\sun}$ for the fast, medium, and slow warm ups, respectively.  This is not a strictly accurate way of determining the relationship between mass at warm-up time, but it leads to much more reasonable masses and takes into account the observational and theoretical work that has been carried out since BM96 was published.

\section{Conclusions}

The disagreement between the disk-like kinematics of the high-mass star-forming region in G35.20-0.74 B and its chemical segregation across its individual cores is not easily explained.  The high cyanide abundances observed toward peak B3 can be reproduced in a fast warm-up, but only with a higher cosmic-ray ionization rate of 1$\times10^{-16}$~s$^{-1}$.  The smallest time period required to reproduce the abundances in source B3 is 3.3~kyr at a gas density of 10$^9$~cm$^{-3}$.  This is a reasonable cosmic-ray ionization rate as evidenced by observations by \citet{Indriolo2015}. The abundances observed in the rest of the disk candidate (B1/B2) can easily be reproduced with a fast warm-up at a gas density of 10$^8$~cm$^{-3}$ and a low rate of cosmic-ray ionization in a very short time period ($\sim$ 1.3~kyr), but can also be reproduced with the higher cosmic-ray ionization rate of 1$\times10^{-16}$~s$^{-1}$ at a gas density of 10$^9$~cm$^{-3}$ in 1.3~kyr.
\par We find that the abundance of ethyl cyanide in particular is maximized in models with a low initial temperature, a high cosmic-ray ionization rate, a long warm-up time, or a lower gas density.  The model is most sensitive to age in the context of a warm-up model (therefore temperature), and to cosmic-ray ionization rate.  It is not sensitive to the initial ice composition (within observed ranges) and not strongly dependent on gas density showing that the warm-up phase determines the composition.  
\par If we assume that the cosmic-ray ionization rate is the same around sources B1/B2 and source B3 at 1$\times10^{-16}$ s$^{-1}$ and the sources have a gas density of 10$^9$~cm$^{-3}$, then the age of sources B1/B2 is 22.3-23.6 kyr while the age of source B3 is 22.7-26 kyr.  This indicates that both these sources began forming within a few thousand years and source B3 is 2000 years older.  Based on an outer disk rotation period between 9700 and 11100~years, this age difference is physically possible.  So we conclude that the detection of CH$_2$CHCN and C$_2$H$_5$CN can indicate a lower limit for the age of a hot core and a nondetection indicates an upper age limit.  This can be useful when observing a potential multiple system at a lower resolution, where if CH$_2$CHCN or C$_2$H$_5$CN is detected toward one part of a source and undetected in others, it indicates a young high-mass system with protostars of different ages. 
\par We have covered a variety of star formation scenarios including a range of gas densities, regions with triggered star formation (starting at temperatures above 10~K), regions with higher and lower cosmic-ray ionization rates, and a range of masses (via warm-up speeds). With this coverage of parameter space we propose that these model results can be used to interpret and predict observations from a variety of embedded high-mass sources and intend to investigate other sources in the future. While we have explored the parameter space in these models comprehensively and noted the trends arising from this analysis, there is still much work to be carried out theoretically and experimentally to understand the gas and ice chemistry of cyanides.  Without this work, our ability to study complex cyanide chemistry will remain hindered.
   
\begin{acknowledgements}
      We would like to thank our referee, Professor Serena Viti, for her constructive comments and quick reading. The PhD project of V. Allen is funded by the Netherlands Organisation for Scientific Research (NWO) and Netherlands Institute for Space Research (SRON). C. Walsh acknowledges NWO (program 639.041.335) and the University of Leeds for financial support.\\
\end{acknowledgements}

\bibliography{bib} 

\begin{appendix} 
\onecolumn
\section{Initial conditions}
\label{appIC}

 \begin{table*}[!ht]
  \centering
  \caption{Full initial conditions (abundance of each species vs. total composition)}
  \begin{tabular}{cccccc}

  \hline\hline
  Species & Initial conditions 1 & Initial conditions 2 & Initial conditions 3 & Initial conditions 4 & Initial conditions 5 \\
  \hline
  H (gas) & 3.525$\times$10$^{-5}$ & 2.560$\times$10$^{-5}$ & 0 & 1.880$\times$10$^{-5}$ & 2.270$\times$10$^{-5}$ \\
  C (gas) & 1.375$\times$10$^{-4}$ & 1.367$\times$10$^{-4}$ & 1.150$\times$10$^{-4}$ & 1.357$\times$10$^{-4}$ & 1.3875$\times$10$^{-4}$ \\
  N (gas) & 7.475$\times$10$^{-5}$ & 7.480$\times$10$^{-5}$ & 7.250$\times$10$^{-5}$ & 7.350$\times$10$^{-5}$ & 7.350$\times$10$^{-5}$ \\
  O (gas) & 3.118$\times$10$^{-4}$ & 3.048$\times$10$^{-4}$ & 2.375$\times$10$^{-4}$ & 3.0385$\times$10$^{-4}$ & 3.029$\times$10$^{-4}$ \\
  H$_2$O (ice) & 5.0$\times$10$^{-6}$ & 1.0$\times$10$^{-5}$ & 5.0$\times$10$^{-5}$ & 1.0$\times$10$^{-5}$ & 1.0$\times$10$^{-5}$ \\
  CO (ice) & 5.0$\times$10$^{-7}$ & 5.0$\times$10$^{-7}$ & 5.0$\times$10$^{-6}$ & 8.0$\times$10$^{-7}$ & 1.7$\times$10$^{-6}$ \\
  CO$_2$ (ice) & 5.0$\times$10$^{-7}$ & 1.5$\times$10$^{-6}$ & 5.0$\times$10$^{-6}$ & 1.3$\times$10$^{-6}$ & 2.3$\times$10$^{-6}$ \\
  NH$_3$ (ice) & 2.5$\times$10$^{-7}$ & 2.0$\times$10$^{-7}$ & 2.5$\times$10$^{-6}$ & 1.5$\times$10$^{-6}$ & 1.5$\times$10$^{-6}$ \\
  CH$_3$OH (ice) & 2.5$\times$10$^{-7}$ & 5.0$\times$10$^{-7}$ & 2.5$\times$10$^{-6}$ & 1.0$\times$10$^{-6}$ & 4.0$\times$10$^{-7}$ \\
  HCOOH (ice) & 5.0$\times$10$^{-7}$ & 5.0$\times$10$^{-7}$ & 5.0$\times$10$^{-6}$ & 7.0$\times$10$^{-7}$ & 1.0$\times$10$^{-7}$ \\
  CH$_4$ (ice) & 2.5$\times$10$^{-7}$ & 1.0$\times$10$^{-7}$ & 2.5$\times$10$^{-6}$ & 1.5$\times$10$^{-7}$ & 1.5$\times$10$^{-7}$ \\
  H$_2$CO (ice) & 5.0$\times$10$^{-7}$ & 2.0$\times$10$^{-7}$ & 5.0$\times$10$^{-6}$ & 3.5$\times$10$^{-7}$ & 2.0$\times$10$^{-7}$ \\
  \hline
  \end{tabular}
  \tablefoot{It is assumed that all available atomic hydrogen is in the form of H$_2$. IC1 is based on a lower limit of the water abundance of 10$^{-5}$ vs. H$_2$ and IC 3 is based on the upper limit of water abundance of 10$^{-4}$ vs. H$_2$. For IC 2, 4, and 5 the water ice abundance is set at 5 $\times$ 10$^{-5}$ vs. H$_2$ and the other ice abundances are calculated from percentages vs. water from observations of ice in star-forming regions \citep{Gibb2004}.  IC2 is based on AFGL 2136, IC4 on W33A, and IC5 on NGC7538 IRS9.}
  \end{table*}

\section{Abundance ranges with errors}
\label{3520errors}

\begin{table*}[htb]
\centering
\setlength{\tabcolsep}{3pt}
\caption{Abundance range observed in \citet{Allen}. Columns 2, 5, and 8 are the best fit abundances; 3, 6, and 9 are the lower limit to the abundances from error calculations; and 4, 7, and 10 are the upper limits to abundances.  CH$_2$CHCN and C$_2$H$_5$CN were not detected in B1 or B2 so their abundances are an upper limit.}
\label{errors}
\begin{tabular}{c|ccc|ccc|ccc}

\hline\hline
                      & \multicolumn{3}{c}{\textbf{A}}                             & \multicolumn{3}{c}{\textbf{B1/B2}}                                        & \multicolumn{3}{c}{\textbf{B3}} \\
\textbf{Species}      & \textbf{Abundance}  & \textbf{Lower}      & \textbf{Upper} & \textbf{Abundance}         & \textbf{Lower}       & \textbf{Upper}       & \textbf{Abundance}   & \textbf{Lower}      & \textbf{Upper} \\
\hline
\textbf{CH$_3$OH}     & 5.04$\times10^{-7}$ & 1.02$\times10^{-7}$ & 1.96$\times10^{-6}$ & 6.68$\times10^{-7}$   & 2.05$\times10^{-7}$  & 4.55$\times10^{-5}$  & 1.36$\times10^{-6}$  & 1.05$\times10^{-6}$ & 1.71$\times10^{-6}$  \\
\textbf{C$_2$H$_5$OH} & 2.97$\times10^{-9}$ & 9.28$\times10^{-10}$ & 3.33$\times10^{-8}$ & 9.42$\times10^{-10}$ & 6.85$\times10^{-10}$ & 1.56$\times10^{-8}$  & 5.15$\times10^{-9}$  & 4.08$\times10^{-9}$ & 6.02$\times10^{-9}$  \\
\textbf{CH$_3$CHO}    & 1.11$\times10^{-9}$ & 3.44$\times10^{-11}$ & 4.17$\times10^{-9}$ & 1.88$\times10^{-9}$  & 1.78$\times10^{-10}$ & 7.11$\times10^{-9}$  & 1.76$\times10^{-9}$  & 1.12$\times10^{-9}$ & 7.56$\times10^{-9}$  \\
\textbf{CH$_3$OCHO}   & 3.38$\times10^{-9}$ & 4.30$\times10^{-10}$ & 4.17$\times10^{-9}$ & 7.02$\times10^{-9}$  & 8.80$\times10^{-10}$ & 1.56$\times10^{-8}$  & 1.10$\times10^{-8}$  & 9.45$\times10^{-9}$ & 1.72$\times10^{-8}$  \\
\textbf{CH$_3$CN}     & 2.97$\times10^{-9}$ & 9.39$\times10^{-10}$ & 1.50$\times10^{-8}$ & 3.93$\times10^{-10}$ & 3.65$\times10^{-10}$ & 5.29$\times10^{-10}$ & 3.10$\times10^{-9}$  & 2.31$\times10^{-9}$ & 4.04$\times10^{-7}$  \\
\textbf{CH$_2$CHCN}   & 5.33$\times10^{-10}$ & 1.44$\times10^{-10}$ & 1.46$\times10^{-9}$ & \multicolumn{3}{c}{Upper limit 2$\times$10$^{-13}$}  & 2.92$\times10^{-10}$ & 1.63$\times10^{-10}$ & 4.00$\times10^{-9}$  \\
\textbf{C$_2$H$_5$CN} & 6.37$\times10^{-10}$ & 1.07$\times10^{-10}$ & 1.25$\times10^{-9}$ & \multicolumn{3}{c}{Upper limit 1$\times$10$^{-13}$}   & 5.15$\times10^{-10}$ & 3.40$\times10^{-10}$ & 7.98$\times10^{-10}$ \\
\textbf{HC$_3$N}      & 5.13$\times10^{-10}$ & 1.42$\times10^{-10}$ & 1.83$\times10^{-9}$ & 5.93$\times10^{-11}$ & 4.79$\times10^{-11}$  & 3.03$\times10^{-9}$ & 3.05$\times10^{-9}$ & 2.40$\times10^{-9}$ & 3.75$\times10^{-9}$ \\
\hline
\end{tabular}
\end{table*}

\section{Comparison with \citet{Garrod2008}}
We compared our model without reaction-diffusion competition to the well-known model in \citet{Garrod2008} and found significant differences.  At all warm-up speeds the difference between our abundances and their reduced model is 1 to 4 orders of magnitude for more complex species, while the abundances of simpler species (H$_2$O, CO, NH$_3$, and CH$_4$) are similar to those in \citet{Garrod2008}. The model abundances from \citet{Garrod2008} cannot reproduce the observed abundances in G35.20 B3, as the fractional abundances of  C$_2$H$_5$OH, CH$_3$OCHO, and CH$_3$CHO are at least one order of magnitude too low.  These authors did not report abundances of CH$_2$CHCN or C$_2$H$_5$CN so that cannot be compared.  There are some notable differences between our model results and those of Garrod. The initial ice composition not the same, although we found that the initial ice composition does not strongly affect the final abundances.  Without knowing their grain surface parameters, that cannot be compared.  Garrod et. al also used a different gas network from us (UMIST versus OSU) and both networks have been updated significantly since 2008.  Most updates to the networks involve updating the binding energies of surface species \citep{Penteado2017}. We also take further steps in varying the cosmic-ray ionization rate and gas densities to investigate the effect of these parameters on the chemical make-up of our modeled sources.

\begin{landscape}

\section{Time ranges}
\label{timeperiods}
Tables and figures showing the time ranges that are required to reproduce observed abundances within errors.

\begin{table*}[htb]
\centering
\setlength{\tabcolsep}{3pt}
\caption{Approximate time period (in years) during which the modeled abundance range matches the observed abundance range for B3. The star symbol indicates that more than one time range fits the observed abundance. The dagger symbol indicates that the observed abundance is not reached by the model (too low).}
\label{B3times}
  {\tiny
\begin{tabular}{c|lll|lll|lll}
\hline\hline
\textbf{B3} & \multicolumn{3}{c}{\textbf{n 10$^7$}} & \multicolumn{3}{c}{\textbf{n 10$^8$}} & \multicolumn{3}{c}{\textbf{n 10$^9$}} \\
 & \textbf{Fast}   & \textbf{Medium} & \textbf{Slow}    & \textbf{Fast}   & \textbf{Medium}  & \textbf{Slow}      & \textbf{Fast}   & \textbf{Medium}   & \textbf{Slow} \\
\hline
CH$_3$OH                       & 22250-22300       & 86000-87000       & 420000-430000      & 22700-22800       & 88000-89000        & 439000               & 23200-23300       & 91000-91500         & 445000-450000      \\
C$_2$H$_5$OH                   & 23400-23600       & 91000-91500       & 430000-440000      & 23800-24000       & 92800-93400        & 440000               & 24500-24700       & 95700-96000         & 455000             \\
CH$_3$CHO                      & 22800-23800       & 85000-91500       & 405000-410000      & 24000-24800       & 86700-89000$\star$ & 415000-420000        & 23200-29000       & 96000-97500         & 435000$\star$       \\
CH$_3$OCHO                     & $\dagger$         & 89000$\star$      & 365000-375000      & 21600-21900       & 90000$\star$       & 470000-475000$\star$ & \textgreater21300 & 99000-101000$\star$ & 485000-495000$\star$ \\
CH$_3$CN                       & \textgreater22000 & \textgreater84000 & \textgreater405000 & \textgreater22500 & \textgreater87000  & \textgreater417000   & \textgreater23300 & \textgreater89700   & \textgreater430000 \\
CH$_2$CHCN                     & \textgreater28500 & 95200-130000      & 440000-475000      & \textgreater29000 & 97600-141000       & 470000-485000        & \textgreater34000 & 100000-145000       & 480000-500000      \\
C$_2$H$_5$CN                   & $\dagger$         & 97500-98500       & 475000             & $\dagger$         & \textgreater102000 & 490000               & $\dagger$         & \textgreater103000  & 500000             \\
HC$_3$N                        & 21600-21800       & 84000-84500       & 400000-405000      & 22100-22300       & 86000-87000        & 418000-420000        & 22700-23000       & 89000-89600         & 430000-435000      \\
\hline
\textbf{Best time period fit}  & no fit      & 84500-97500       & 375000-475000      & no fit      & 87000-102000       & 420000-490000        & no fit      & 89600-103000        & 435000-500000      \\
\textbf{Temperatures (K)}      & 95-158            & 93-116            & 78-120             & 96-164            & 98-133             & 96-127               & 105-223           & 104-134             & 102-132            \\
Limiters                       & HC$_3$N-C$_2$H$_3$CN       & HC$_3$N-C$_2$H$_5$CN       & CH$_3$OCHO-C$_2$H$_5$CN     & CH$_3$OCHO-CH$_2$CHCN    & HC$_3$N-C$_2$H$_5$CN        & HC$_3$N-C$_2$H$_5$CN          & HC$_3$N-CH$_2$CHCN       & HC$_3$N-C$_2$H$_5$CN         & HC$_3$N-C$_2$H$_5$CN       \\
\hline
\end{tabular}
}
\end{table*}

\begin{table*}[htb]
\centering

\setlength{\tabcolsep}{3pt}
\caption{As Table~\ref{B3times} for B1/B2.  Because vinyl and ethyl cyanide were not detected, the rows for CH$_2$CHCN and C$_2$H$_5$CN are the model abundances for the time period. The star symbol indicates that more than one time range fits the observed abundance.}
\label{B1B2times}
  {\tiny
\begin{tabular}{c|lll|lll|lll}
\hline\hline
\textbf{B1/B2} & \multicolumn{3}{c}{\textbf{n 10$^7$}} & \multicolumn{3}{c}{\textbf{n 10$^8$}} & \multicolumn{3}{c}{\textbf{n 10$^9$}} \\
 & \textbf{Fast}   & \textbf{Medium} & \textbf{Slow}    & \textbf{Fast}   & \textbf{Medium}  & \textbf{Slow}      & \textbf{Fast}   & \textbf{Medium}   & \textbf{Slow} \\
\hline
CH$_3$OH                     & \textgreater21800     & \textgreater85000            & \textgreater415000    & \textgreater22200     & \textgreater87000     & \textgreater425000    & \textgreater22800     & \textgreater89000            & \textgreater435000 \\
C$_2$H$_5$OH                 & \textgreater22900     & 90000-93000                  & 425000-445000         & \textgreater23300     & 91000-95000           & 435000-450000         & 24000-26000           & 93800-97500                  & 440000-465000 \\
CH$_3$CHO                    & 21700-23800           & 81500-91300                  & 330000-410000         & 23000-24800           & 83000-95000           & 405000-425000         & 22600-29000           & 94500-97500                  & 460000-480000 \\
CH$_3$OCHO                   & \textgreater25000     & 75500-78500$\star$           & 360000-375000         & 21900-22000$\star$    & 90000-90600$\star$    & 455000-475000$\star$  & 21000-21500           & 91000-92000*                 & 445000$\star$ \\
CH$_3$CN                     & 19000-20000           & 74500-75500                  & 360000-365000         & 22100-22200           & 85000-85500           & 407000-410000         & 22600-22900           & 87600-88000                  & 418000-420000 \\
CH$_2$CHCN                   & 10$^{-12}$–10$^{-11}$ & 10$^{-12}$–10$^{-11}$        & 10$^{-11}$–10$^{-10}$ & 10$^{-13}$            & 10$^{-12}$            & 10$^{-12}$            & 10$^{-14}$–10$^{-13}$ & 6$\times$10$^{-14}$–10$^{-13}$ & 8$\times$10$^{-13 }$–5$\times$10$^{-12}$ \\
C$_2$H$_5$CN                 & 10$^{-14}$–10$^{-12}$ & 10$^{-14}$–5$\times$10$^{-13}$ & 10$^{-13}$–10$^{-12}$ & 10$^{-15}$–10$^{-14}$ & 10$^{-15}$–10$^{-14}$ & 10$^{-14}$            & 10$^{-17}$–10$^{-15}$ & 7$\times$10$^{-16}$            & 10$^{-14}$   \\
HC$_3$N                      & 20000-21800           & 79500-84000                  & 365000-405000         & 21000-22300           & 82400-86800           & 400000-420000         & 21600-22900           & 84900-89500                  & 410000-432000 \\
\hline
\textbf{Best time period fit}& 20000-25000           & 75500-90000                  & 365000-425000         & 22000-23300           & 85500-91000           & 410000-455000         & 21500-24000           & 88000-94500                  & 420000-460000 \\
\textbf{Temperatures (K)}    & 81-123                & 76-105                       & 75-98                 & 96-107                & 95-107                & 92-111                & 92-114                & 100-114                      & 96-113 \\
Limiters                     & CH$_3$CN-CH$_3$OCHO         & CH$_3$CN-C$_2$H$_5$OH                 & CH$_3$CN-C$_2$H$_5$OH          & CH$_3$OCHO-C$_2$H$_5$OH        & CH$_3$CN-C$_2$H$_5$OH          & CH$_3$CN-CH$_3$OCHO         & CH$_3$OCHO-C$_2$H$_5$OH        & CH$_3$CN-CH$_3$CHO                 & CH$_3$CN-CH$_3$CHO  \\
\hline
\end{tabular}
}
\end{table*}

\begin{table*}[htb]
\centering

\setlength{\tabcolsep}{3pt}
\caption{As Table~\ref{B3times} for A. The star symbol indicates that more than one time fits the observed abundance.}
\label{Atimes}
  {\tiny
\begin{tabular}{c|lll|lll|lll}
\hline\hline
\textbf{A} & \multicolumn{3}{c}{\textbf{n 10$^7$}} & \multicolumn{3}{c}{\textbf{n 10$^8$}} & \multicolumn{3}{c}{\textbf{n 10$^9$}} \\
 & \textbf{Fast}   & \textbf{Medium} & \textbf{Slow}    & \textbf{Fast}   & \textbf{Medium}  & \textbf{Slow}      & \textbf{Fast}   & \textbf{Medium}   & \textbf{Slow} \\
\hline
CH$_3$OH                       & 21700-22300       & 84000-87000        & 415000-430000        & 22000-22800        & 86500-89000        & 423000-438000   & 22700-23200       & 88500-92000        & 438000-460000  \\
C$_2$H$_5$OH                   & \textgreater23000 & 90000-96200        & 430000-440000$\star$ & \textgreater23500  & 92000-96500        & 439000-452000   & \textgreater24100 & 94000-99500        & 450000-465000  \\
CH$_3$CHO                      & 13800-23400       & 52000-90000        & 260000-410000        & 22300-24500        & 81500-94300        & 395000-420000   & 22000-23500       & 83000-97000        & 450000-475000  \\
CH$_3$OCHO                     & \textgreater24500 & 89000-90500$\star$ & 350000-365000        & 21700-22100$\star$ & 93000-95300$\star$ & 460000-465000   & 20500-21000       & 91000-92000$\star$ & 470000-485000$\star$ \\
CH$_3$CN                       & 21300-23000       & 80000-87000        & 370000-410000        & 22400-23800        & 86000-89000        & 415000-425000   & 23000-24700       & 89000-92000        & 425000-435000  \\
CH$_2$CHCN                     & \textgreater28500 & 95000-105000       & 435000-465000        & \textgreater28500  & 97500-110000       & 470000-480000   & \textgreater32000 & 99500-110000       & 480000-490000  \\
C$_2$H$_5$CN                   & \textgreater25200 & 97300-99000        & 472000-480000        & \textgreater29000  & \textgreater105000 & 490000          & not reproduced    & \textgreater103000 & 500000         \\
HC$_3$N                        & 21000-21600       & 81500-85000        & 380000-400000        & 21400-22100        & 83500-86000        & 405000-415000   & 22000-22700       & 86000-88700        & 418000-430000  \\
\hline
\textbf{Best time period fit}  & 21600-28500       & 85000-97300        & 365000-472000        & 22100-28500        & 86000-105000       & 415000-490000   & no fit      & 88700-103000       & 430000-500000  \\
\textbf{Temperatures (K)}      & 93-158            & 94-121             & 75-119               & 97-158             & 96-139             & 94-127          & 88-198            & 102-134            & 100-132        \\
Limiters                       & HC$_3$N CH$_2$CHCN       & HC$_3$N C$_2$H$_5$CN        & CH$_3$OCHO C$_2$H$_5$CN       & HC$_3$N C$_2$H$_5$CN        & HC$_3$N C$_2$H$_5$CN        & HC$_3$N C$_2$H$_5$CN     & CH$_3$OCHO CH$_2$CHCN    & HC$_3$N C$_2$H$_5$CN        & HC$_3$N C$_2$H$_5$CN    \\
\hline
\end{tabular}
}
\end{table*}

   \begin{figure}[h]
   \centering
      \includegraphics[width=\hsize]{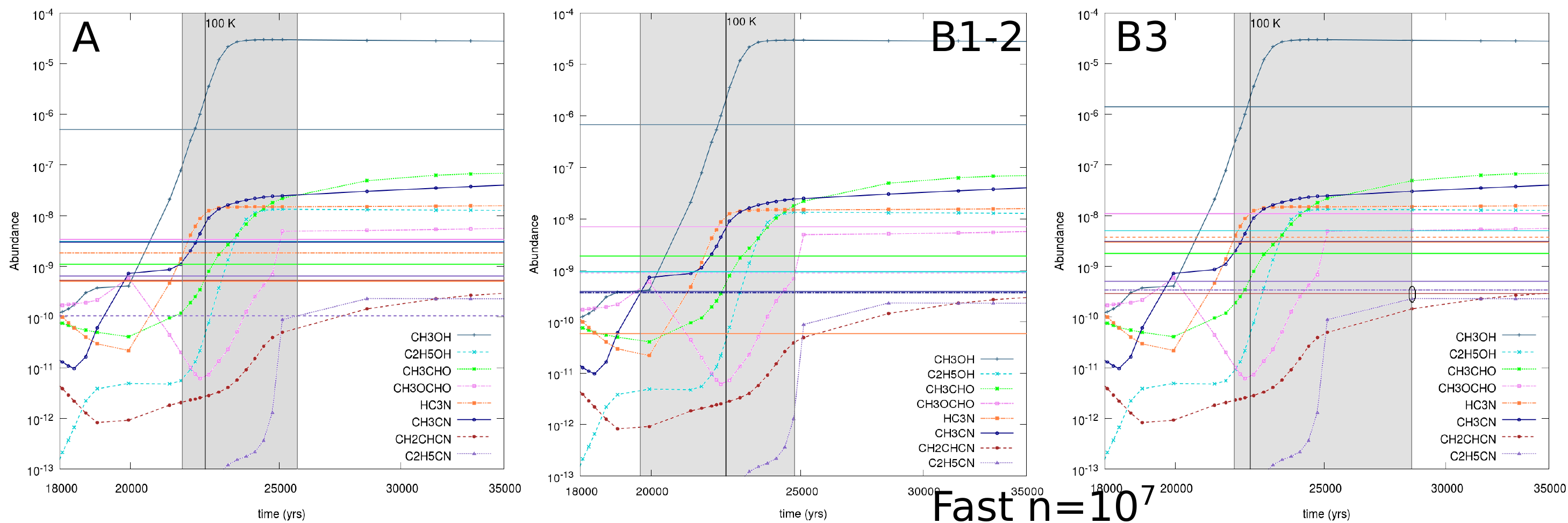}
      \caption{Abundances vs. H$_2$ for CH$_3$OH, C$_2$H$_5$OH, CH$_3$CHO, CH$_3$OCHO, HC$_3$N, CH$_3$CN, CH$_2$CHCN, and C$_2$H$_5$CN using IC 5 with a density of 10$^7$ cm$^{-3}$ and a fast warm-up time of 50 kyr are shown for G35.20 A (left), B1/B2 (middle), and B3 (right).  The time period shown is only a part of the modeled time, from 18000-35000 yr.  The time range in which all abundances can be reproduced with an error of 1 order of magnitude are shaded in gray. The abundance of C$_2$H$_5$CN in B3 is not reproduced so a small black ellipse shows the gap between the lower abundance limit and the modeled abundance.}
   \end{figure}

   \begin{figure}[h]
   \centering
      \includegraphics[width=\hsize]{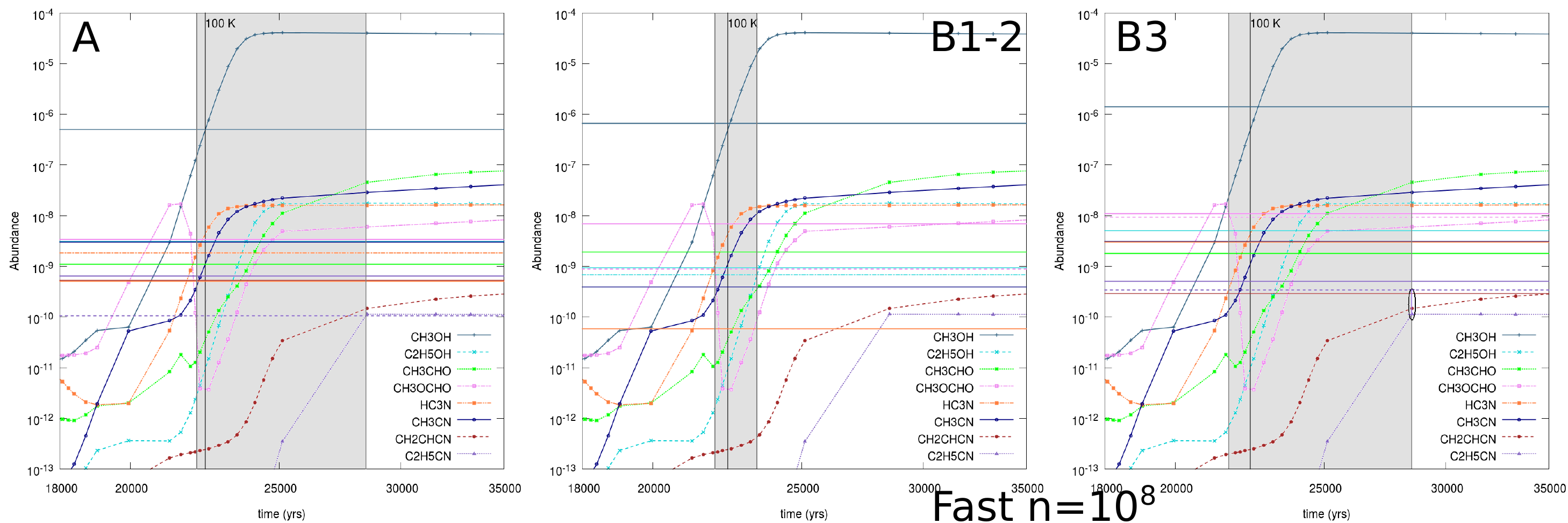}
      \caption{Abundances vs. H$_2$ for CH$_3$OH, C$_2$H$_5$OH, CH$_3$CHO, CH$_3$OCHO, HC$_3$N, CH$_3$CN, CH$_2$CHCN, and C$_2$H$_5$CN using IC 5 with a density of 10$^8$ cm$^{-3}$ and a fast warm-up time of 50 kyr are shown for G35.20 A (left), B1/B2 (middle), and B3 (right).  The time period shown is only a part of the modeled time, from 18000-35000 yr.  The time range in which all abundances can be reproduced with an error of 1 order of magnitude are shaded in gray.  The abundance of C$_2$H$_5$CN in B3 is not reproduced so a small black ellipse shows the gap between the lower abundance limit and the modeled abundance.}
   \end{figure}

   \begin{figure}[h]
   \centering
      \includegraphics[width=\hsize]{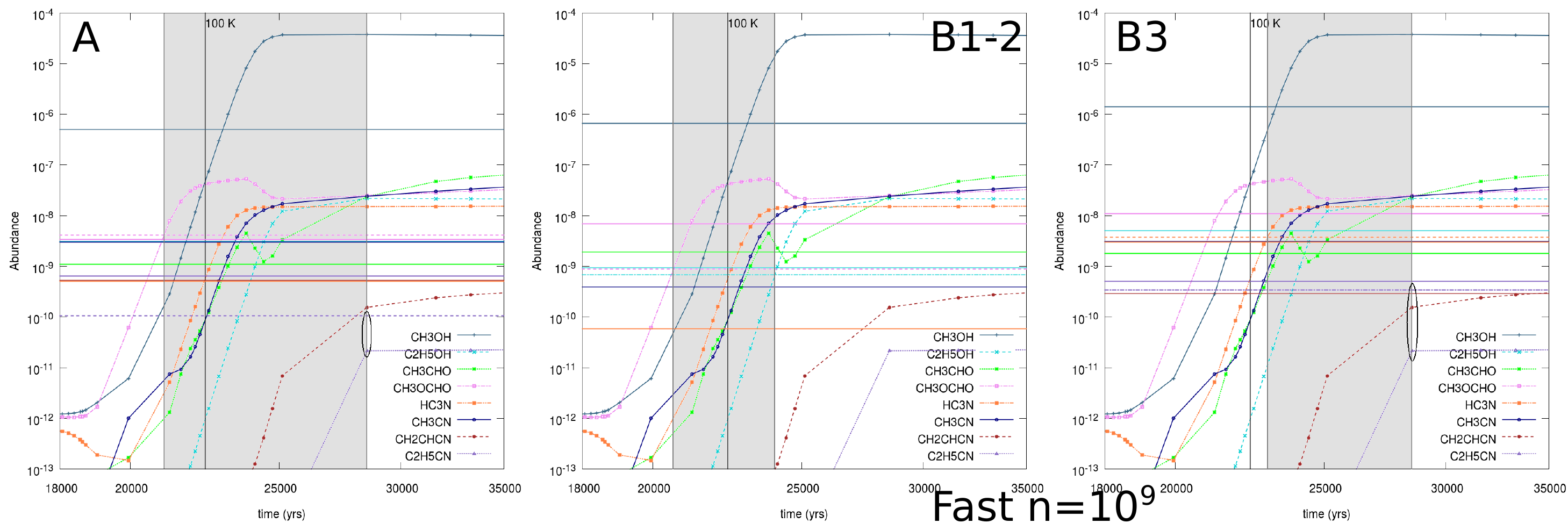}
      \caption{Abundances vs. H$_2$ for CH$_3$OH, C$_2$H$_5$OH, CH$_3$CHO, CH$_3$OCHO, HC$_3$N, CH$_3$CN, CH$_2$CHCN, and C$_2$H$_5$CN using IC 5 with a density of 10$^9$ cm$^{-3}$ and a fast warm-up time of 50 kyr are shown for G35.20 A (left), B1/B2 (middle), and B3 (right).  The time period shown is only a part of the modeled time, from 18000-35000 yr.  The time range in which all abundances can be reproduced with an error of 1 order of magnitude are shaded in gray. The abundance of C$_2$H$_5$CN in A and B3 is not reproduced so a small black ellipse shows the gap between the lower abundance limit and the modeled abundance.}
   \end{figure}

   \begin{figure}[h]
   \centering
      \includegraphics[width=\hsize]{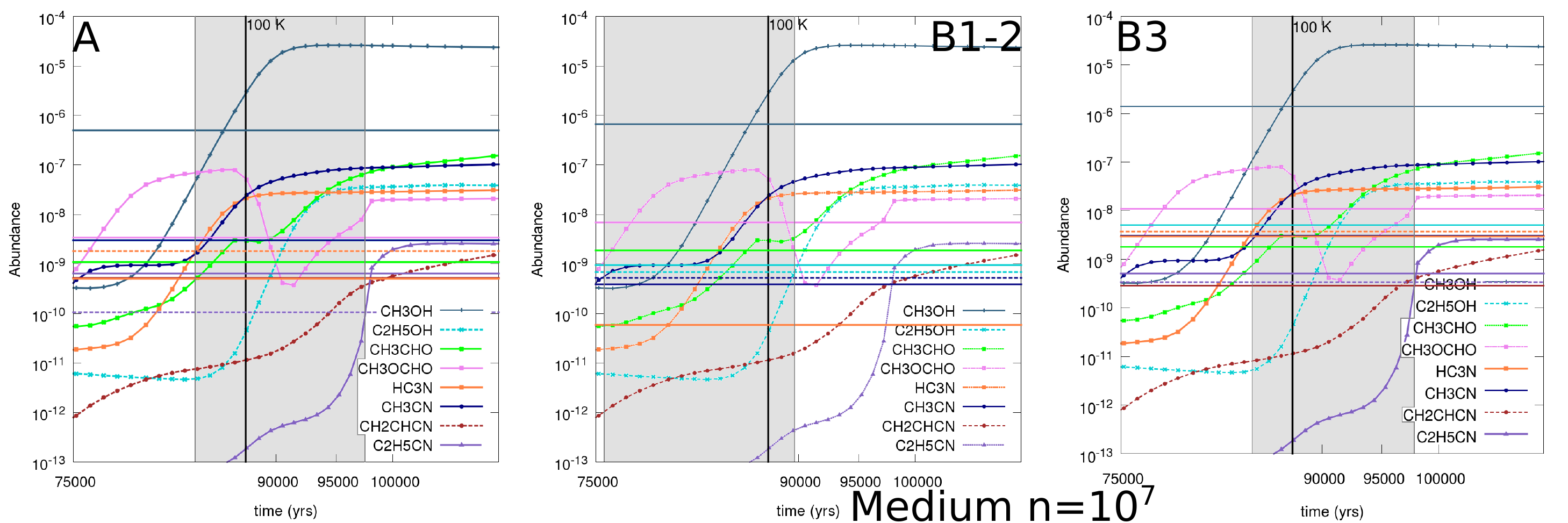}
      \caption{Abundances vs. H$_2$ for CH$_3$OH, C$_2$H$_5$OH, CH$_3$CHO, CH$_3$OCHO, HC$_3$N, CH$_3$CN, CH$_2$CHCN, and C$_2$H$_5$CN using IC 5 with a density of 10$^7$ cm$^{-3}$ and a medium warm-up time of 200 kyr are shown for G35.20 A (left), B1/B2 (middle), and B3 (right).  The time period shown is only a part of the modeled time, from 70-120 kyr for A and B3 and 30-105 kyr for B1/B2.  The time range in which all abundances can be reproduced with an error of 1 order of magnitude are shaded in gray.}
   \end{figure}

   \begin{figure}[h]
   \centering
      \includegraphics[width=\hsize]{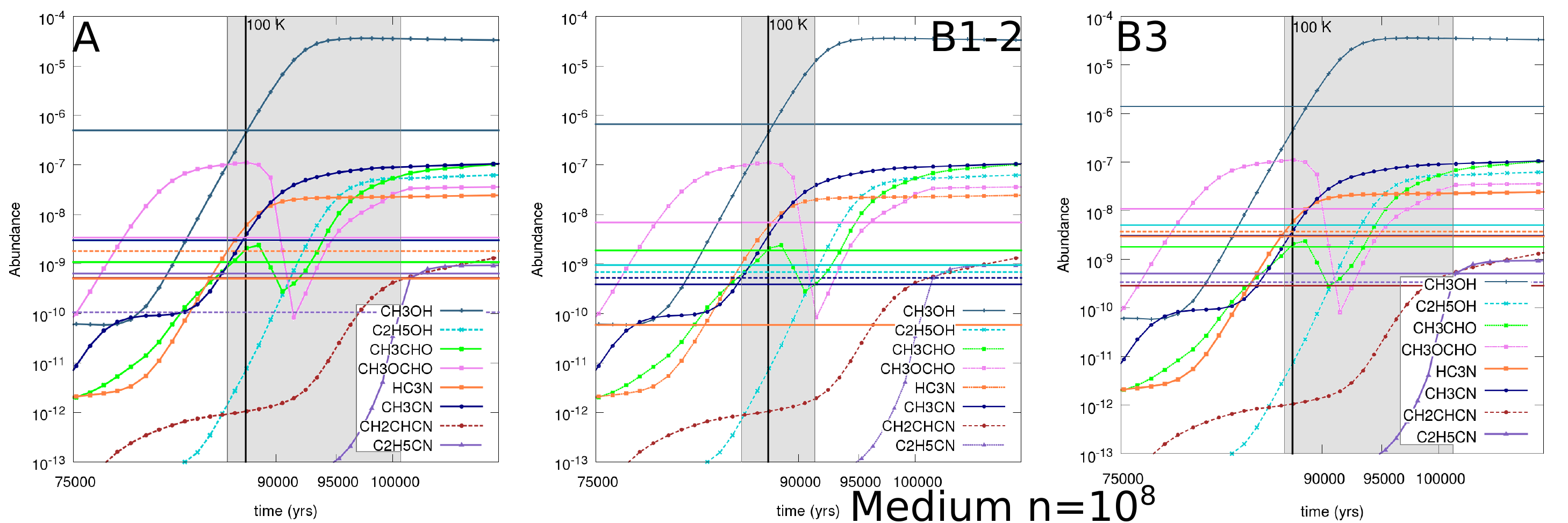}
      \caption{Abundances vs. H$_2$ for CH$_3$OH, C$_2$H$_5$OH, CH$_3$CHO, CH$_3$OCHO, HC$_3$N, CH$_3$CN, CH$_2$CHCN, and C$_2$H$_5$CN using IC 5 with a density of 10$^8$ cm$^{-3}$ and a medium warm-up time of 200 kyr are shown for G35.20 A (left), B1/B2 (middle), and B3 (right).  The time period shown is only a part of the modeled time, from 70-120 kyr for A and B3 and 30-105 kyr for B1/B2.  The time range in which all abundances can be reproduced with an error of 1 order of magnitude are shaded in gray. The abundance of C$_2$H$_5$CN in B3 is not reproduced.}
   \end{figure}

    \begin{figure}[h]
   \centering
      \includegraphics[width=\hsize]{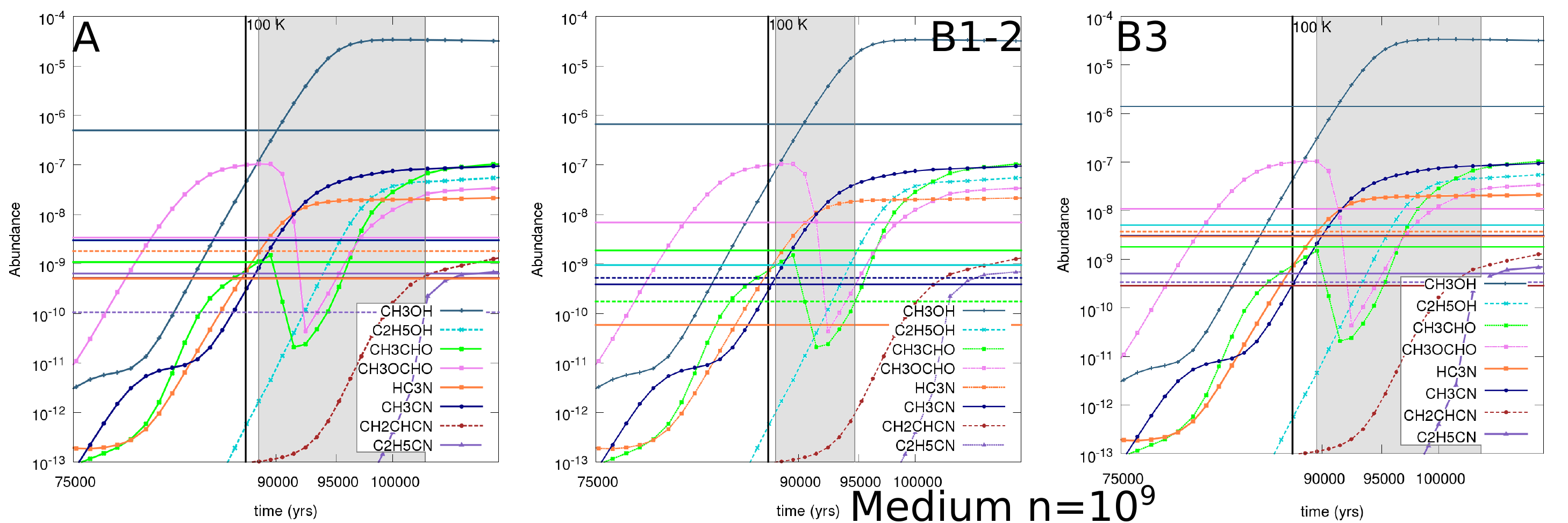}
      \caption{Abundances vs. H$_2$ for CH$_3$OH, C$_2$H$_5$OH, CH$_3$CHO, CH$_3$OCHO, HC$_3$N, CH$_3$CN, CH$_2$CHCN, and C$_2$H$_5$CN using IC 5 with a density of 10$^9$ cm$^{-3}$ and a medium warm-up time of 200 kyr are shown for G35.20 A (left), B1/B2 (middle), and B3 (right).  The time period shown is only a part of the modeled time, from 70-120 kyr for A and B3 and 30-105 kyr for B1/B2.  The time range in which all abundances can be reproduced with an error of 1 order of magnitude are shaded in gray. The abundance of C$_2$H$_5$CN in B3 is not reproduced.}
   \end{figure}

   \begin{figure}[h]
   \centering
      \includegraphics[width=\hsize]{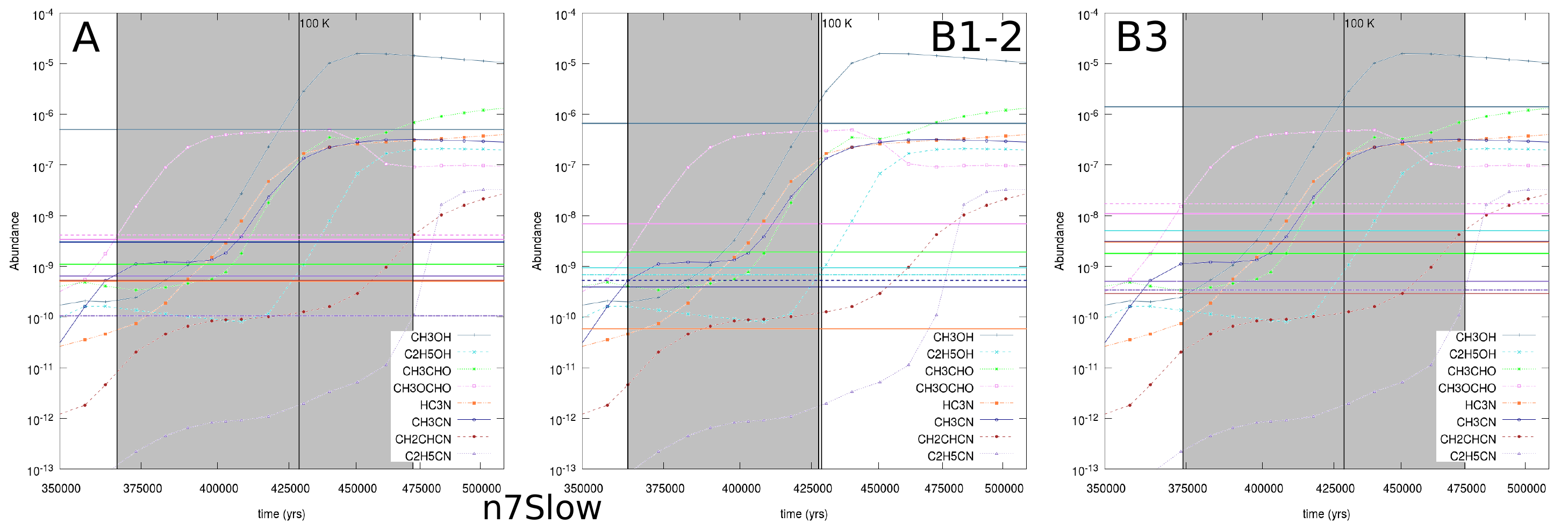}
      \caption{Abundances vs. H$_2$ for CH$_3$OH, C$_2$H$_5$OH, CH$_3$CHO, CH$_3$OCHO, HC$_3$N, CH$_3$CN, CH$_2$CHCN, and C$_2$H$_5$CN using IC 5 with a density of 10$^7$ cm$^{-3}$ and a slow warm-up time of 1 Myr are shown for G35.20 A (left), B1/B2 (middle), and B3 (right).  The time period shown is only a part of the modeled time, from 350-510 kyr.  The time range in which all abundances can be reproduced with an error of 1 order of magnitude are shaded in gray.}
   \end{figure}

   \begin{figure}[h]
   \centering
      \includegraphics[width=\hsize]{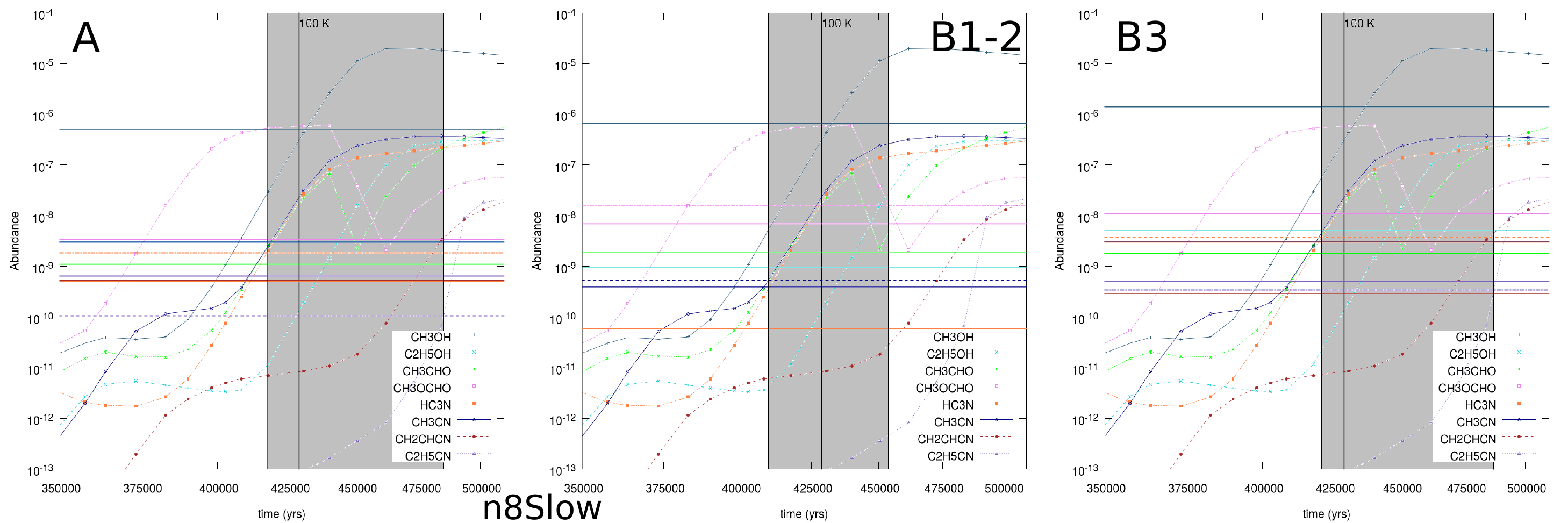}
      \caption{Abundances vs. H$_2$ for CH$_3$OH, C$_2$H$_5$OH, CH$_3$CHO, CH$_3$OCHO, HC$_3$N, CH$_3$CN, CH$_2$CHCN, and C$_2$H$_5$CN using IC 5 with a density of 10$^7$ cm$^{-3}$ and a slow warm-up time of 1 Myr are shown for G35.20 A (left), B1/B2 (middle), and B3 (right).  The time period shown is only a part of the modeled time, from 350-510 kyr.  The time range in which all abundances can be reproduced with an error of 1 order of magnitude are shaded in gray.}
   \end{figure}

   \begin{figure}[h]
   \centering
      \includegraphics[width=\hsize]{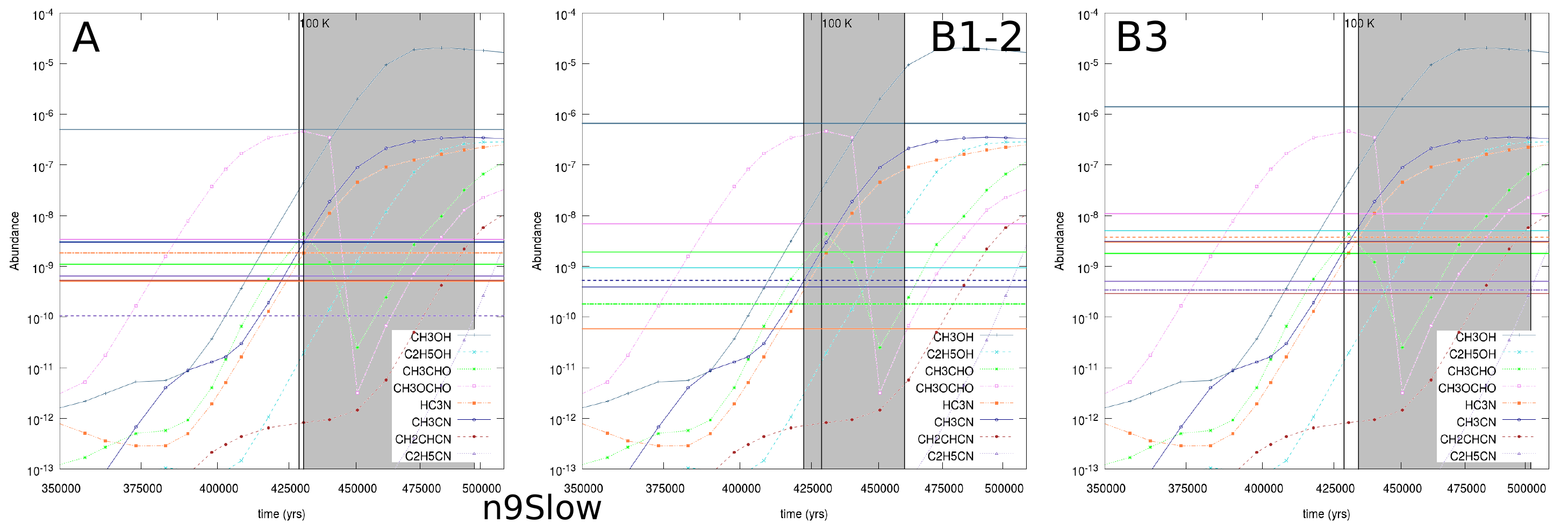}
      \caption{Abundances vs. H$_2$ for CH$_3$OH, C$_2$H$_5$OH, CH$_3$CHO, CH$_3$OCHO, HC$_3$N, CH$_3$CN, CH$_2$CHCN, and C$_2$H$_5$CN using IC 5 with a density of 10$^9$ cm$^{-3}$ and a slow warm-up time of 1 Myr are shown for G35.20 A (left), B1/B2 (middle), and B3 (right).  The time period shown is only a part of the modeled time, from 350-510 kyr.  The time range in which all abundances can be reproduced with an error of 1 order of magnitude are shaded in gray.}
   \end{figure}
\end{landscape}

\end{appendix}

\end{document}